\documentclass[useAMS,usenatbib]{mn2e}
\usepackage{epsf}
\usepackage{epsfig}
\usepackage{subfigure}
\usepackage{epstopdf}
\usepackage{amsmath}
\usepackage{xcolor}

\newcommand{\refsim}{\textsc{ref}}
\newcommand{\agna}{\textsc{agn 8.0}}
\newcommand{\agnb}{\textsc{agn 8.5}}
\newcommand{\agnc}{\textsc{agn 8.7}}
\newcommand{\calsim}{\textsc{bahamas}}
\newcommand{\nocool}{\textsc{nocool}}
\newcommand{\nua}{\textsc{nu 0.00}}
\newcommand{\nub}{\textsc{nu 0.06}}
\newcommand{\nuc}{\textsc{nu 0.12}}
\newcommand{\nud}{\textsc{nu 0.24}}
\newcommand{\nue}{\textsc{nu 0.48}}
\newcommand{\nuadm}{\textsc{nu 0.00 dm}}

\newcommand{\wmapa}{\textit{WMAP}7}   
\newcommand{\wmapb}{\textit{WMAP}9}
\newcommand{\dmonly}{\textsc{dmonly}}
\newcommand{\deltaTheat}{$\Delta T_{\rm heat}$}
\newcommand{\conc}{$c_{200}$}
\newcommand{\mconc}{$\langle c_{200}\rangle$}
\newcommand{\fdbk}{baryonic feedback}
\newcommand{\nufs}{neutrino free-streaming}
\newcommand{\mnu}{$\sum M_{\nu}$}                     
\newcommand{\mc}{$c(M)$}
\newcommand{\sims}{BAHAMAS and cosmo-OWLS}

\title[Baryons, neutrinos, and large-scale structure]{The separate and combined effects of baryon physics and neutrino free-streaming on large-scale structure}                       
                                                                      
\author[B.~Mummery et~al.]{Benjamin O.~Mummery$^1$\thanks{E-mail: b.o.mummery@2010.ljmu.ac.uk}, Ian G. McCarthy$^1$\thanks{Email: i.g.mccarthy@ljmu.ac.uk}, Simeon Bird$^2$, Joop Schaye$^3$\\
  $^{1}$Astrophysics Research Institute, Liverpool John Moores University, 146 Brownlow Hill, Liverpool L3 5RF\\
$^2$Department of Physics and Astronomy, Johns Hopkins University, 3400 N.~Charles Street, Baltimore, MD 21218, USA\\
$^3$Leiden Observatory, Leiden University, P. O. Box 9513, 2300 RA Leiden, the Netherlands}

\begin{document}
  \date{Accepted ... Received ...}                                      
  \pagerange{\pageref{firstpage}--\pageref{lastpage}} \pubyear{2014}

  \maketitle                                                          
  \label{firstpage}                                                   

  \begin{abstract}                                                    

We use the cosmo-OWLS and \calsim~suites of cosmological hydrodynamical simulations to explore the separate and combined effects of baryon physics (particularly feedback from active galactic nuclei, AGN) and free-streaming of massive neutrinos on large-scale structure. We focus on five diagnostics: i) the halo mass function; ii) halo mass density profiles; iii) the halo mass$-$concentration relation; iv) the clustering of haloes; and v) the clustering of matter; and we explore the extent to which the effects of baryon physics and neutrino free-streaming can be treated independently. Consistent with previous studies, we find that both AGN feedback and neutrino free-streaming suppress the total matter power spectrum, although their scale and redshift dependencies differ significantly. The inclusion of AGN feedback can significantly reduce the masses of groups and clusters, and increase their scale radii.  These effects lead to a decrease in the amplitude of the mass$-$concentration relation and an increase in the halo autocorrelation function at fixed mass. Neutrinos also lower the masses of groups and clusters while having no significant effect on the shape of their density profiles (thus also affecting the mass$-$concentration relation and halo clustering in a qualitatively similar way to feedback).  We show that, with only a small number of exceptions, the combined effects of baryon physics and neutrino free-streaming on all five diagnostics can be estimated to typically better than a few percent accuracy by treating these processes independently (i.e., by multiplying their separate effects). 

  \end{abstract}                                                      
                                                                                      
                  
\section{Introduction}                        
\label{sec:intro}

Recent simulation-based work has shown that various physical processes associated with galaxy formation (e.g., radiative cooling, star formation, and feedback processes) can significantly affect not only the predicted distribution of baryons, but also that of the underlying dark matter component.  For example, it has been shown that both the total matter power spectrum (e.g., \citealt{vanDaalen2011,Schneider2015}) and the halo mass function (e.g., \citealt{Sawala2013,Cui2014,Velliscig2014,Cusworth2014,Schaller2015}) can be affected at the tens of percent level relative to that predicted by a standard gravity-only dark matter simulation.  If these effects are ignored, they are expected to lead to significant biases in cosmological parameters inferred by comparing predicted and observed aspects of large-scale structure \textcolor{black}{(}LSS\textcolor{black}{)} (e.g., \citealt{Semboloni2011,Eifler2015,HarnoisDeraps2015}).

However, galaxy formation is not the only process that affects the resultant distribution of LSS.  Recently, there has been a resurgence in interest in the effects of massive neutrinos.  This resurgence has been driven by the apparent tension in the observed abundance of massive clusters compared to that predicted when a {\it Planck} cosmology based on the primary CMB is adopted (e.g., \citealt{Planck2013,Planck2015}), in conjunction with similar tensions between the {\it Planck} primary CMB constraints and those derived from tomographic analysis of cosmic shear data \citep{Heymans2013,Hildebrandt2017}.  It has been argued that massive neutrinos can potentially reconcile th\textcolor{black}{is} tension (e.g., \citealt{Wyman2014,Battye2014}), although this remains controversial (e.g., \citealt{MacCrann2015}).  Regardless of whether neutrinos resolve the tension, atmospheric and solar oscillation experiments have found that the three active species of neutrinos have a summed mass of at least 0.06 eV (0.1 eV) when adopting a normal (inverted) hierarchy \citep{Lesgourgues2006}.  The fact that neutrinos have appreciable mass and will act as a form of hot dark matter that will resist significant gravitational collapse (due to free streaming motion), implies that they will affect the predicted LSS.  Whether these effects are minor or dominant in comparison to those due to galaxy formation is presently unclear and depends on the (relatively poorly constrained) absolute mass scale of the neutrinos and the efficiencies of relevant feedback processes.

Given that both baryon physics and massive neutrinos likely play a role in the formation of LSS in the Universe, it is important to consider their combined effect and whether it amounts to more (or less) than `the sum of its parts'.  That is, to what extent is there cross-talk between the baryon physics and neutrinos?  Do they suppress or enhance each other's effects on LSS, or can they be treated separately?

The aim of the present study is to address these questions by means of direct numerical simulation.  That is, we consider the effects baryon physics and massive neutrinos both separately and in combination, using the recent cosmo-OWLS \citep{LeBrun2014,McCarthy2014} and \calsim~\citep{McCarthy2017} suites of cosmological hydrodynamical simulations.  The two suites are complementary, in that cosmo-OWLS varies the implemented subgrid physics for stellar and AGN feedback at fixed cosmology (with massless neutrinos), while \calsim~varies the neutrino mass for a fixed (calibrated) feedback model.  We further complement these simulations with reference dark matter-only simulations (both with massless and massive neutrinos).  This combination of compl\textcolor{black}{e}mentary simulations provides an unprecedented opportunity to examine the effects of both baryon feedback and the free-streaming of massive neutrinos not simply in isolation\textcolor{black}{,} but also capturing their combined effects on large-scale structure.

We examine five different ways of characterising LSS: i) the halo mass function; ii) total mass density profiles in bins of halo mass; iii) the mass$-$concentration relation; iv) the spatial clustering of haloes (characterised by the 3D 2-point autocorrelation function); and v) the clustering of matter (characterised by the total matter power spectrum).  We demonstrate that both feedback and neutrino free-streaming can have considerable effects on these aspects of LSS and that, to typically better than a few percent accuracy, their combined effects can be estimated by treating these processes independently (i.e., by multiplying their separate effects).  

The paper is organised as follows.  In Section 2 we present a brief summary of the cosmo-OWLS and \calsim~simulations.  In Section 3 we examine the effects of baryon physics and neutrinos on the abundance of haloes.  In Section 4 we examine their effects on the total mass density profiles and the mass$-$concentration relation.  In Sections 5 and 6 we explore how the spatial clustering of haloes and matter (respectively) are affected.  Finally, in Section 7 we summarize and discuss our findings.
                                                                   
\section{ Simulations}                       
\label{sec:sims}
We use the cosmo-OWLS \citep{LeBrun2014,McCarthy2014} and \calsim~\citep{McCarthy2017} suites of cosmological hydrodynamical simulations, both of which are descendants of the OverWhelmingly Large Simulations (OWLS) project \citep{Schaye2010}.  As already noted, the two suites are complementary, in that cosmo-OWLS varies the implemented subgrid physics for stellar and AGN feedback at fixed cosmology, while \calsim~varies the cosmology for a fixed (calibrated) feedback model.  Below we provide a brief overview of the simulations, but we refer the reader to \citet{LeBrun2014} and \citet{McCarthy2017} for further details of the simulations and comparisons with the observed properties of present-day galaxy groups and clusters.  

Table \ref{tab:sims} provides a summary of the included subgrid physics and the model parameter values for the various cosmo-OWLS and \calsim~runs we use.

\begin{table*} 
\caption{\label{tab:sims} Included subgrid physics and model parameter values for the \mbox{cosmo-OWLS} and \mbox{BAHAMAS} runs used here.\ \  (1) Simulation name; (2) inclusion of photoionizing ultra-violet and X-ray backgrounds according to \citet{Haardt2001}; (3) inclusion of radiative cooling and star formation; (4) for runs with star formation, $V_w$ is the velocity kick (in km/s) adopted in the stellar feedback (with a fixed mass-loading of $2$); (5) inclusion of AGN feedback; (6) $\Delta T_{\rm heat}$ is the temperature by which gas is heated by AGN feedback; (7) $n_{\rm heat}$ is the number of gas particles heated by AGN feedback; (8) inclusion of massive neutrinos; (9) the summed mass of neutrinos (assuming a normal hierarchy of neutrino masses); (10) dark matter particle mass; (11) initial baryon particle mass.  A more detailed discussion of these parameters can be found in Section \ref{sec:sims}.}                                                 
\begin{tabular}{l *{10}{c}}                                                                 \hline
(1) & (2) & (3) & (4) & (5) & (6) & (7) & (8) & (9) & (10) & (11) \\
   Simulation & UV/X-ray   & Cooling and    & $V_{w}$ & AGN      & $\Delta T_{\rm heat}$ & $n_{\rm heat}$ & $\nu$ & $M_{\nu}$     & $M_{DM}$                   & $M_{bar,init}$			\\
              & background & star formation & [km/s]  & feedback & [K]               &            &       & [eV]		      & [$10^9 {\rm M_{\odot}}/h$] & [$10^8 {\rm M_{\odot}}/h$]	\\
  \hline
  {\bf cosmo-OWLS} \\          
  \hline
  \nocool & Yes	& No  & -  	& No  & -    	     & -	& No  & -     & 3.75 	 & 7.54\\
  \refsim & Yes & Yes & -  	& No  & -          & -	& No  & -     & 3.75 	 & 7.54\\ 
  \agna   & Yes & Yes & 600 & Yes & $10^{8.0}$ & 1  & No  & - 	& 3.75 	 & 7.54\\
  \agnb   & Yes & Yes & 600	& Yes & $10^{8.5}$ & 1  & No  & -    & 3.75 	 & 7.54\\
  \agnc   & Yes & Yes & 600	& Yes & $10^{8.7}$ & 1 	& No  & -     & 3.75 	 & 7.54\\
  \dmonly & No  & No  & -   & No 	& -          & -  & No  & - 	& 4.50   & - \\
  \hline
  {\bf BAHAMAS} \\
  \hline
  \nua  	& Yes & Yes & 300 & Yes & $10^{7.8}$ & 20 & Yes & massless & 3.85 & 7.66 \\
  \nub 	  & Yes	& Yes & 300 & Yes & $10^{7.8}$ & 20 & Yes & 0.06 	   & 3.83 & 7.66 \\
  \nuc 		& Yes & Yes & 300 & Yes & $10^{7.8}$ & 20 & Yes & 0.12 	   & 3.81 & 7.66 \\
  \nud 		& Yes & Yes & 300 & Yes & $10^{7.8}$ & 20 & Yes & 0.24 	   & 3.77 & 7.66 \\
  \nue 		& Yes & Yes & 300	& Yes & $10^{7.8}$ & 20 & Yes & 0.48 	   & 3.68 & 7.66 \\
  \nua~DM & No  & No  & -  	& No 	& -          & -  & Yes & massless & 4.62 & - \\
  \nub~DM & No	& No  & -  	& No  & -          & -  & Yes & 0.06 	   & 4.61 & - \\
  \nuc~DM	& No  & No  & -  	& No  & -          & -  & Yes & 0.12 	   & 4.58 & - \\
  \nud~DM	& No 	& No  & - 	& No 	& -          & -  & Yes & 0.24 	   & 4.53 & - \\
  \nue~DM	& No 	& No 	& -	  & No 	& -          & -  & Yes & 0.48 	   & 4.44 & - \\
  \hline
\end{tabular}
\end{table*} 

\subsection{cosmo-OWLS}

The cosmo-OWLS suite of cosmological hydrodynamical simulations consists of $400 \ {\rm Mpc}/h$ comoving on a side, periodic box simulations containing $2 \times 1024^3$ particles.  The simulations adopt a cosmology based on the maximum likelihood parameter values obtained from the analysis of \wmapa~data \citep{Komatsu2011}; i.e., \{$\Omega_{m}$, $\Omega_{b}$, $\Omega_{\Lambda}$, $\sigma_{8}$, $n_{s}$, $h$\} = \{0.272, 0.0455, 0.728, 0.81, 0.967, 0.704\}.  The algorithm of \citet{Eisenstein1999} was used to compute the transfer function and {\small N-GenIC}\footnote{http://www.mpa-garching.mpg.de/gadget/} (developed by V.~Springel) was used to make the initial conditions, at a starting redshift of $z=127$.  The dark matter and (initial) baryon particle masses are $\approx3.75\times10^{9}~h^{-1}~\textrm{M}_{\odot}$ and $\approx7.54\times10^{8}~h^{-1}~\textrm{M}_{\odot}$, respectively.  The gravitational softening is fixed to $4~h^{-1}$ kpc (in physical coordinates below $z=3$ and in comoving coordinates at higher redshifts).

The simulations were carried out with a version of the Lagrangian TreePM-SPH code \textsc{gadget3} \citep[last described in][]{Springel2005a}, which was modified to include new subgrid physics as part of the OWLS project.  We use all five baryon physics models presented in Le Brun et al. (all of which adopted identical initial conditions), along with a corresponding dark matter only run.  The models are:   
\begin{itemize}
 \item \dmonly: A dissipationless \textcolor{black}{``}dark matter-only\textcolor{black}{''} simulation.
 \item \nocool :  A standard non-radiative model; i.e., inclusion of baryons and hydrodynamics but no subgrid modules for radiative cooling, star formation, etc.
 \item \refsim: In addition to the inclusion of baryons and hydrodynamics, this model includes prescriptions for element-by-element radiative cooling \citep*{Wiersma2009a}, star formation \citep{Schaye2008}, stellar evolution, mass loss and chemical enrichment \citep{Wiersma2009b} from Type II and Ia supernovae and Asymptotic Giant Branch stars, and kinetic stellar feedback \citep{DallaVecchia2008}.
 \item \agna, \agnb, and \agnc: In addition to the physics included in the \refsim~model, these models include a prescription for supermassive black hole (BH) growth and AGN feedback \citep{Springel2005b,Booth2009}.  In brief, an on-the-fly friends-of-friends (FOF) algorithm, with a linking length of 0.2 times the mean interparticle separation, is run during the simulation and any FOF haloes identified with at least 100 dark matter particles that do not already contain a BH `sink' particle are seeded with one, with an initial mass of 0.001 times the initial gas particle mass.  BH particles then grow in mass via mergers with other BH particles and through gas accretion, as described in \citet{Booth2009}.  In terms of feedback, the BHs accumulate the feedback energy in a reservoir until they are able to heat neighbouring gas particles by a pre-determined amount $\Delta T_{\rm heat}$.  cosmo-OWLS uses 1.5 per cent of the rest-mass energy of the gas which is accreted on to the supermassive black holes for the AGN feedback, which results in a good match to the normalisation of the black hole scaling relations (\citealt{Booth2009,LeBrun2014}), independent\textcolor{black}{ly} of the exact value of $\Delta T_{\rm heat}$. The three AGN models differ only by their value of $\Delta T_{\rm heat}$, which is the most important parameter of the feedback model in terms of the gas-phase properties of the resulting group and cluster population (\citealt{LeBrun2014,McCarthy2017}).  It is set to $\Delta T_{\rm heat}=10^{8.0}$ K for \agna, $\Delta T_{\rm heat}=10^{8.5}$ K for \agnb, and $\Delta T_{\rm heat}=10^{8.7}$ K for \agnc. Note that since the same quantity of gas is being heated in these models, more time is required for the black holes to accrete a sufficient amount of gas to heat the adjacent gas to a higher temperature.  Therefore, increased heating temperatures lead to more episodic and violent feedback events.
\end{itemize}

We note that the range of cosmo-OWLS models considered here is somewhat extreme, in that the models that neglect AGN feedback (\refsim\ and \nocool) have significantly higher total baryon fractions than observed for local X-ray-bright galaxy groups (e.g., \citealt{Sun2009}), while the AGN model with the most extreme feedback, \agnc, yields galaxy groups with gas fractions that are considerably lower than observed.  (The more moderate AGN feedback models, \agna\ and \agnb, skirt the upper and lower bounds of the observed trend between hot gas mass and halo mass for X-ray-bright galaxy groups; see \citealt{LeBrun2014}.)  However, two important caveats are that: i) the role of observational selection effects is not well understood for galaxy groups (e.g., current observations cannot rule out the existence of a population of virialized groups which are X-ray faint and may have lower gas fractions); and ii) there are too few observational constraints on high-redshift systems to judge whether or not the various models are realistic at earlier times.  Bearing these caveats in mind, we have elected to explore the trends using the ensemble of cosmo-OWLS models.

A resolution study for cosmo-OWLS can be found in Appendix A of \citet{LeBrun2014}, where it is demonstrated that the gas and stellar mass fractions of the simulated groups and clusters are reasonably well converged (i.e., change by only a few percent over an increase in mass resolution of a factor of 8).

\subsection{BAHAMAS}

In common with cosmo-OWLS, the \calsim~suite presented in \citet{McCarthy2017} consists of $400 \ {\rm Mpc}/h$ comoving on a side, periodic box simulations containing $2 \times 1024^3$ particles.  In the present study, we use a subset of the \calsim~suite whose initial conditions are based on the updated maximum-likelihood cosmological parameters derived from the \wmapb~data \citep{Hinshaw2013}; i.e., \{$\Omega_{m}$, $\Omega_{b}$, $\Omega_{\Lambda}$, $\sigma_{8}$, $n_{s}$, $h$\} = \{0.2793, 0.0463, 0.7207, 0.821, 0.972, 0.700\}.

We also use a massive neutrino extension of \calsim~recently completed by McCarthy et al. (in prep).  Specifically, using the semi-linear algorithm of \citet{Bird2013}, McCarthy et al.\ have run massive neutrino versions of the \wmapb~cosmology for several different choices of the total summed neutrino mass, $M_\nu$, ranging from the minimum mass implied by neutrino oscillation experiments of $\approx 0.06$ eV \citep{Lesgourgues2006} up to $0.48$ eV, in factors of two.  When implementing massive neutrinos, all other cosmological parameters are held fixed apart from $\sigma_8$ and the matter density in cold dark matter, which was decreased slightly to maintain a flat model (i.e., so that $\Omega_{\rm b}+\Omega_{\rm cdm}+\Omega_\nu+\Omega_\Lambda=1$).  The parameter $\sigma_8$ characterises the amplitude of linear theory $z=0$ matter density fluctuations on $8 h^{-1}$ Mpc scales.  Instead of holding this number fixed, the amplitude, $A_s$, of the density fluctuations at the epoch of recombination (as inferred by \wmapb~data assuming massless neutrinos) is held fixed, in order to retain agreement with the observed CMB angular power spectrum.  Other strategies for implementing neutrinos are also possible (e.g., decreasing $\Omega_\Lambda$ instead of $\Omega_{\rm cdm}$) but McCarthy et al.~have found with small test simulations that the precise choice of what is held fixed (apart from the power spectrum amplitude) does not have a large effect on the local cluster population.  Most important is the value of $\Omega_\nu$, which is related to $M_\nu$ via $\Omega_\nu = M_\nu / (93.14 \ {\rm eV} \ h^2)$ \citep{Lesgourgues2006} and ranges from 0.0013 to 0.0105 for our choices of summed neutrino mass.  For completeness, the runs with $M_\nu =$ 0.0, 0.06, 0.12, 0.24, and 0.48 eV have $\sigma_8$ = 0.821, 0.813, 0.799, 0.766, \textcolor{black}{and} 0.705, respectively.

For both the runs with and without massive neutrinos, the Boltzmann code {\small CAMB}\footnote{http://camb.info/} (\citealt{Lewis2000}; April 2014 version) was used to compute the transfer functions and a modified version of {\small N-GenIC} to create the initial conditions, at a starting redshift of $z=127$.  {\small N-GenIC} has been modified by S.\ Bird to include second-order Lagrangian Perturbation Theory corrections and support for massive neutrinos\footnote{https://github.com/sbird/S-GenIC}.  Note that in producing the initial conditions for \calsim\textcolor{black}{,} we use the separate transfer functions computed by {\small CAMB} for each individual component (baryons, neutrinos, and CDM), whereas in cosmo-OWLS (and indeed \textcolor{black}{in} most existing cosmological hydro simulations) the baryons and CDM adopt the same transfer function, corresponding to the total mass-weighted function.  Note also that we use the same random phases for each of the simulations, implying that intercomparisons between the different runs are not subject to cosmic variance complications.

The \calsim~runs used here have dark matter and (initial) baryon particle masses for a \wmapb~massless neutrino cosmology of $\approx3.85\times10^{9}~h^{-1}~\textrm{M}_{\odot}$  and $\approx7.66\times10^{8}~h^{-1}~\textrm{M}_{\odot}$, respectively.  (The particle masses differ only slightly from this when massive neutrinos are included; see Table \ref{tab:sims}.)  The gravitational softening of the runs presented is fixed to $4~h^{-1}$ kpc, as in cosmo-OWLS.

\begin{figure*}
 \includegraphics[width=0.80\textwidth]{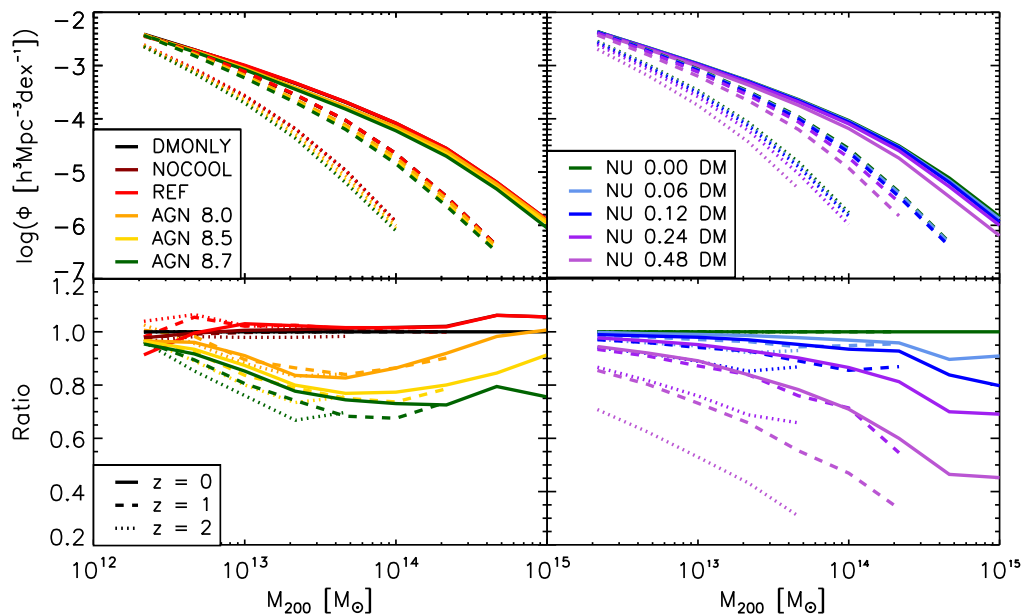}
 \caption{Halo mass functions (HMFs) for the different baryon physics runs (in the absence of neutrino physics) from cosmo-OWLS (left) and the different collisionless massive neutrino runs from \calsim\ (right).  Colours denote the various runs (see the legend and Table \ref{tab:sims}), while the different linestyles denote different redshifts.   In the bottom left panel, the cosmo-OWLS HMFs have been normalised by the \dmonly\ case, whereas in the bottom right panel the HMFs have been normalised by the massless neutrino case.  Suppression of the HMF due to AGN feedback (orange, yellow and green, left) is important at intermediate (group) masses but becomes less important at high halo masses, where it begins to converge towards the \dmonly\ case (with the mass scale where the convergence occurs depending on the AGN heating temperature \deltaTheat).  The suppression due to feedback is only a weak function of redshift.  In the collisionless neutrino simulations (right panel), the suppression is strongest for the highest mass haloes and, in contrast to the effects of feedback, exhibits a strong redshift dependence.  }
 \label{fig:a_hmf_1}  
\end{figure*}

The \calsim~runs were carried out with the same version of the \textsc{gadget3} code that was used in \mbox{(cosmo-)OWLS}.  As noted above, to perform runs with massive neutrinos included, McCarthy et al.~used the semi-linear algorithm developed by \citet{Bird2013} (see also \citealt{Bond1980,Ma1995,Brandbyge2008,Brandbyge2009,Bird2012}), implemented in the \textsc{gadget3} code.  Schematically, the semi-linear code computes neutrino perturbations on the fly at every time step using a linear perturbation integrator {\it sourced from the non-linear baryons+CDM potential}, adding the result to the total gravitational force.  Because the neutrino power is calculated at every time step, the dynamical response\textcolor{black}{s} of the neutrinos to the baryons+CDM and of the baryons+CDM to the neutrinos \textcolor{black}{are} mutually and self-consistently included.  Note that because the integrator uses perturbation theory\textcolor{black}{,} the method does not account for the non-linear response of \textcolor{black}{the} neutrino component to itself.  However, this limitation has negligible consequences for our purposes, as only a very small fraction of the neutrinos (with lower \textcolor{black}{velocities} than typical) are expected to collapse and the neutrinos as a whole constitute only a small fraction of the total matter density\footnote{We have explicitly tested this by comparing the predicted mass density profiles of simulated groups and clusters using the semi-linear algorithm with that predicted using a particle-based treatment of the massive neutrinos (e.g., \citealt{Viel2010,Bird2012}), for simulations with CDM and neutrinos but no baryons.  The resulting mass profiles \textcolor{black}{typically} agree to better than two percent accuracy over the full range of radii resolved in the simulations.}.

\textcolor{black}{I}n addition to neutrinos, the various \calsim~runs (with or without massive neutrinos) also include the effects of radiation when computing the background expansion rate.  We find that this leads to a few percent reduction in the amplitude of the present-day linear matter power spectrum compared to a simulation that only considers the evolution of dark matter and dark energy in the background expansion rate.

\calsim~differs significantly from cosmo-OWLS in terms of its approach to the choice of parameter values for the subgrid feedback.  In particular, \citet{McCarthy2017} explicitly calibrated the stellar and AGN feedback models to reproduce the observed present-day galaxy stellar mass function and the amplitude of the hot gas mass$-$halo mass relation of groups and clusters respectively, as determined by X-ray observations.  By calibrating to these observables, the simulated groups and clusters are guaranteed to have the correct baryon content in a global sense.  The associated back reaction of the baryons on the total matter distribution should therefore also be broadly correct.  \citet{McCarthy2017} have shown that the \calsim~simulations reproduce an unprecedentedly wide range of properties of massive systems, including the various observed mappings between galaxies, hot gas, total mass, and black holes.

A resolution study for \calsim~is presented in Appendix C of \citet{McCarthy2017}, where it is demonstrated that the gas and stellar mass fractions are reasonably well converged (to better than $\approx10\%$ in the case of a strong test, and to $\approx2\%$ in the case of a weak test, using the terminology of \citealt{Schaye2015}) over the range of halo masses that we consider in the present study. 

\section{Halo Abundances} 

\begin{figure}
 \includegraphics[width=0.995\columnwidth]{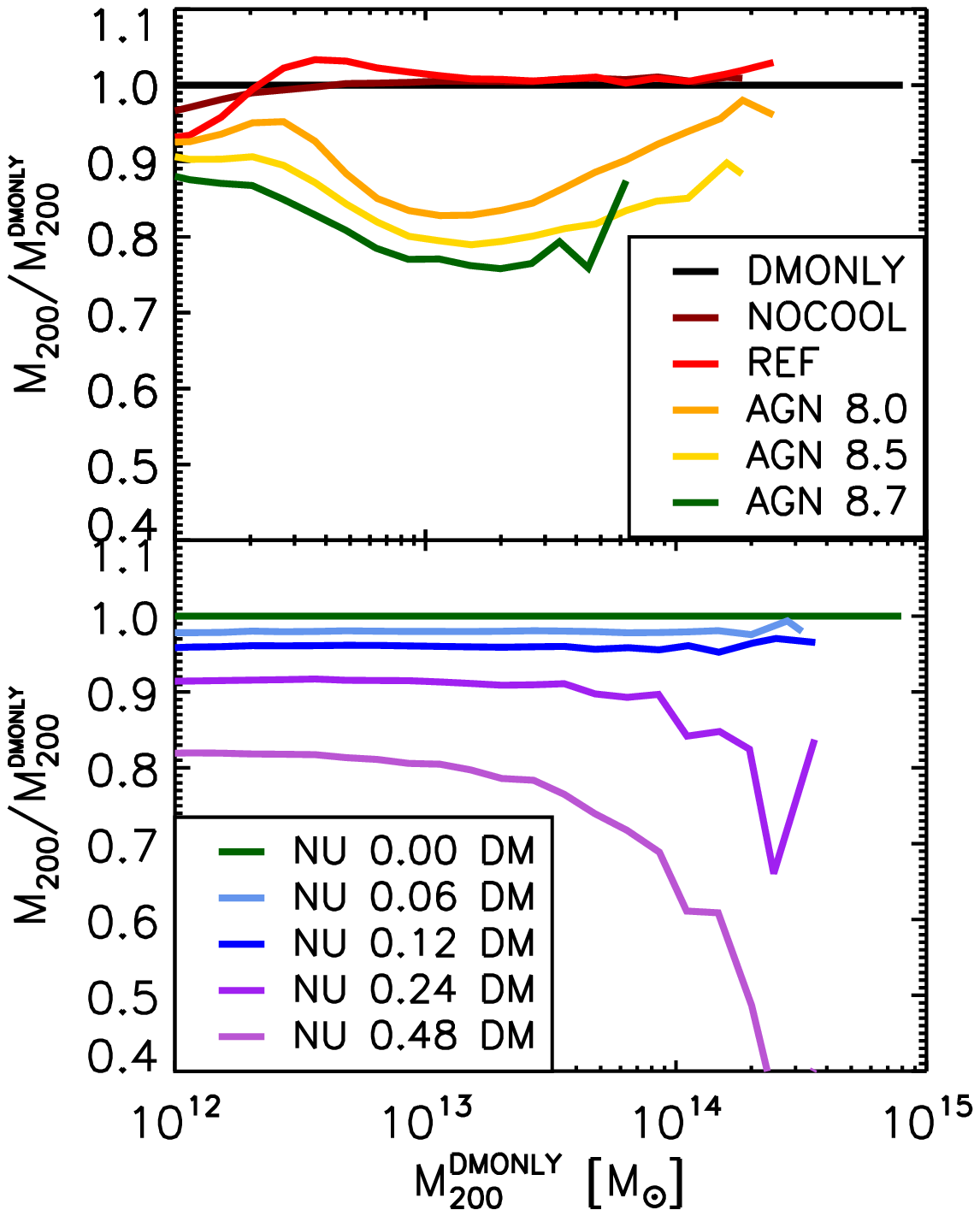}
 \caption{\label{fig:mass_change_2}
  The fractional change in the halo mass, relative to the DM-only, massless neutrino case, arising from the inclusion of \fdbk\ (left panel) and \nufs\ (right panel) at $z=0$. As in Fig. \ref{fig:a_hmf_1}, colours denote the various \sims\ runs as detailed in Table \ref{tab:sims}.  {\it Top:} AGN feedback can reduce the mass of a halo by up to $\textcolor{black}{\approx}20\%$ at group masses, but tends towards the \dmonly\ case at higher masses with the mass scale for convergence depending on the choice of AGN heating temperature \deltaTheat.  {\it Bottom:} The free-streaming of massive neutrinos in collisionless simulations reduces halo masses to a similar degree at group masses, but is of increasing importance at higher masses.  These effects drive those seen in the HMFs (Fig. \ref{fig:a_hmf_1}).    
}
\end{figure}

\subsection{Halo Mass Functions} \label{sec:HMF}

We begin by examining the effects of baryonic physics and the inclusion of massive neutrinos, both separately and in combination, on the halo mass function (HMF).  We define the halo mass function, $\Phi$, as the number of haloes with mass $M_{200,{\rm crit}}$ per comoving cubic Mpc per logarithmic unit mass; i.e., $\Phi \equiv dn/d\log_{10}(M_{200,{\rm crit}})$, where $M_{200,{\rm crit}}$ is the mass contained within a radius that encloses a mean density of 200 times the universal critical density at that redshift.  Haloes are identified using a standard FOF algorithm run on the dark matter distribution, with a linking length of 0.2 in units of the mean interparticle separation.  We use the \textsc{subfind} algorithm \citep{Springel2001,Dolag2009} to calculate the spherical overdensity mass $M_{200,{\rm crit}}$ (i.e., the mass contained within the radius that encloses a mean density that is $200 \times$ the critical density at that redshift). These spheres are centred on the position of the main subhalo's particle \textcolor{black}{with the minimum gravitational potential.}

In the left panel of Fig.~\ref{fig:a_hmf_1} we plot the HMFs for the various baryon physics runs (without neutrinos) from cosmo-OWLS.  For runs that lack feedback from AGN, the HMF, as expected, largely follows that of the \dmonly\ case, at least for the range of halo mass in which we are interested here.  However, when one includes feedback from AGN the situation changes significantly -- gas is ejected from the high-redshift progenitors of groups and clusters \citep{McCarthy2011} leading to a significant suppression (of up to $\approx20 - 30\%$) of the HMF at masses of $\sim 10^{13-14}{\rm {\rm M_{\odot}}}$ in $M_{200,{\rm crit}}$, in agreement with that previously reported by \citet{Velliscig2014}, who analysed a subset of the cosmo-OWLS runs (see also \citealt{Cui2014,Cusworth2014}).  Note that the reduction in the baryonic mass also leads to a shallowing of the gravitational potential well. This causes the dark matter distribution to expand outwards becoming less densely concentrated, and also results in a reduction in the accretion rate onto the main progenitor (e.g., \citealt{Sawala2013,Velliscig2014}), which is why the $M_{200,{\rm crit}}$ masses of individual haloes can be reduced by somewhat more than the universal baryon fraction of $\Omega_b/\Omega_m$.  

The deeper potential wells of higher-mass ($M_{200,{\rm crit}} \ga 10^{14} {\rm {\rm M_{\odot}}}$) systems are able to retain a larger fraction of their baryons.  Consequently, the behaviour in the HMF tends back towards the \dmonly\ case at the highest masses.  The precise mass scale where the AGN runs converge towards the \dmonly\ case depends on the adopted heating temperature, with higher heating temperatures increasing this mass scale, as one would anticipate based on the strong dependence of the baryon fraction on the AGN heating temperature reported previously by \citet{LeBrun2014} and \citet{McCarthy2017}. At lower masses (below $10^{13} {\rm M_{\odot}}$), the trend\textcolor{black}{s} for the AGN cases also tend back towards the \dmonly\ case.  This is due to inefficient accretion onto the black holes. The precise location of this convergence in \textcolor{black}{the} simulations, however, is sensitive both to the initial mass of black hole sink particles and the halo mass at which they are seeded.

In contrast, neutrino free-streaming (right panel of Fig.~\ref{fig:a_hmf_1}) preferentially suppresses the high-mass end of the HMF (see also \citealt{Costanzi2013}).  This is due to the fact that the effect of the free-streaming of massive neutrinos on the linear matter power spectrum grows with time, and appears in the clustering statistics from the collapse redshift of the cluster.  Consequently, more massive objects, which collapse later in CDM-based cosmologies, are more strongly affected by neutrino free-streaming.  The strength of this suppression also varies strongly as a function of the summed neutrino mass, with higher values leading to a stronger reduction of the HMF.   Interestingly, while the suppression due to \fdbk\ (left panel of Fig.~\ref{fig:a_hmf_1}) is only weakly dependent on redshift, the massive neutrino runs show stronger evolution with redshift.

While Fig. \ref{fig:a_hmf_1} shows the change in number density at fixed halo mass, we \textcolor{black}{also} want to explicitly examine the change in halo mass at fixed number density, since this is more physical (i.e., feedback and neutrinos do not affect the abundance of haloes, they alter their masses).  In order to determine the effects of \fdbk\ and \nufs\ on individual haloes, we co\textcolor{black}{n}struct a matched set of haloes across all simulation runs.  Haloes are matched using the unique particle IDs for the dark matter particles. For each particle assigned to a halo in the DM-only, massless neutrino case, the particle with the matching ID is identified in each of the other simulations. \textcolor{black}{In each case, t}he halo in each case containing the highest number of identified particles is selected as the match. This method finds matches for $\approx 83\%$ and $\approx 90\%$ of haloes in the cosmo-OWLS and \calsim\ cases respectively, for haloes in the range $12 \le \log(M_{200,{\rm crit}}/{\rm M_\odot}) \le 15$ where the $M_{200,{\rm crit}}$ value under consideration is that of the halo in the DM-only, massless neutrino case.

We show in Fig. \ref{fig:mass_change_2} the fractional change in halo mass as a function of the mass in the dark matter only, massless neutrino case.  Unsurprisingly, the behaviour of the alteration to halo mass arising from \fdbk\ and \nufs\ is almost identical to their effects on the HMF (Fig. \ref{fig:a_hmf_1}).

Henceforth, when using the matched sample of haloes we use for each halo the values of $M_{200}$ and $r_{200}$ that correspond to the matching halo in the DM-only, massless neutrino case

\begin{figure}
 \includegraphics[width=0.995\columnwidth]{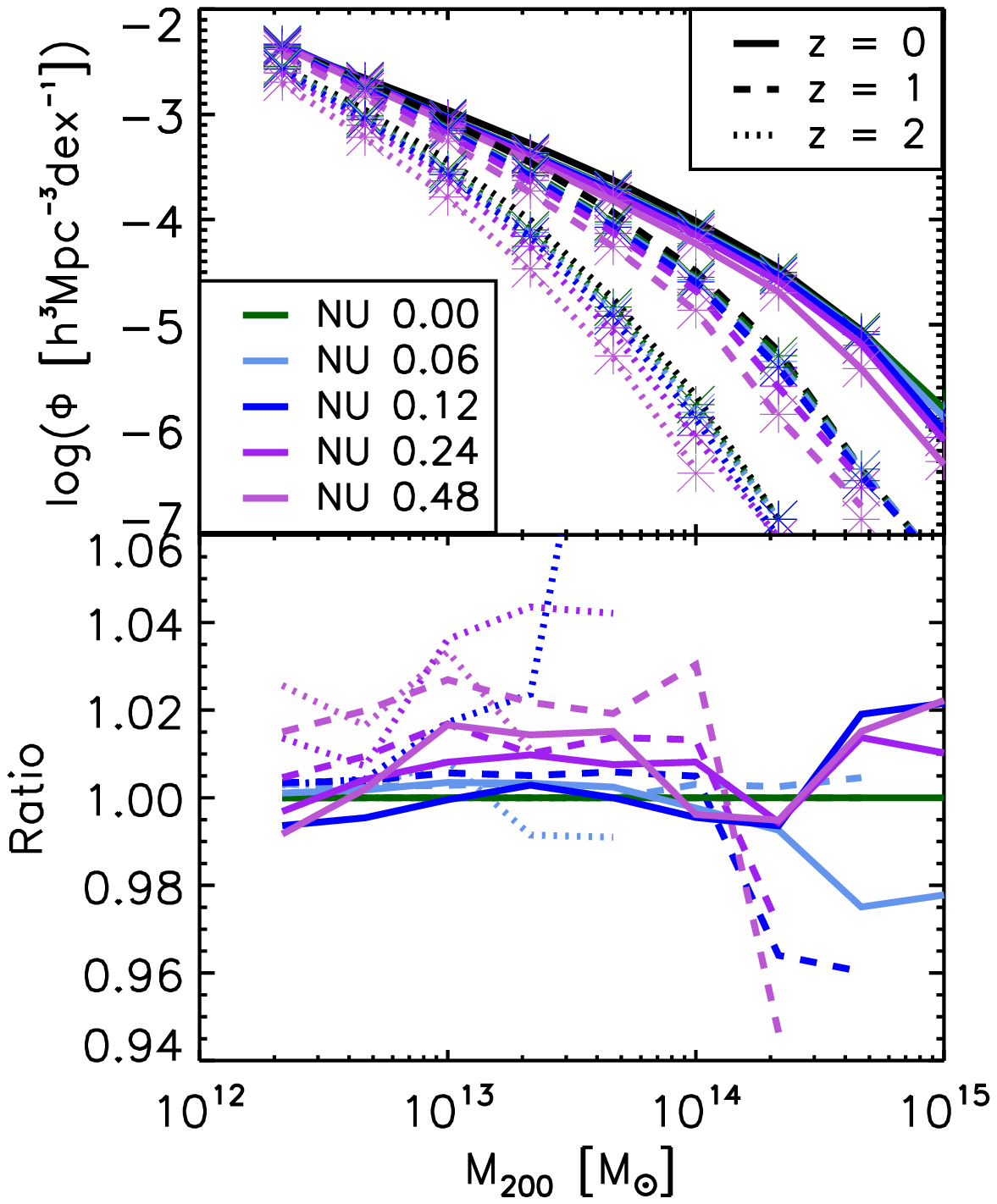}
 \caption{Comparison of the halo mass functions arising when simultaneously simulating \fdbk\ and \nufs, and those calculated by multiplying the separate effects of \fdbk\ in the absence of neutrinos and the effects of \nufs\ in the absence of baryons. {\it Top:} Curves display the HMFs arising when simultan\textcolor{black}{e}ously simulating \nufs\ and \fdbk. The multiplicative calculations are displayed by crosses. In both cases, colours c\textcolor{black}{o}rrespond to different values for the summed neutrino mass while solid, dashed and dotted curves display the results at redshifts of 0, 1 and 2 respectively. {\it Bottom:} The ratio of each simultan\textcolor{black}{e}ous simulation to the corresponding multiplicative prediction. The two cases agree to within a few percent accuracy over the full range of halo masses and redshifts that we have examined. }
 \label{fig:a_hmf_2}
\end{figure}

\subsubsection{Separability}
While the effects of baryon physics and neutrino free-streaming have individually been investigated in a number of previous studies (although generally with much poorer statistics), their combined effect has not been examined.  In particular, it is unclear to what extent the baryonic effects (particularly gas expulsion from AGN feedback) and neutrino free-streaming are separable.  That is, can these processes be treated independently, or do they amplify (or perhaps suppress) each other?  

To answer this question, the top panel of Fig.~\ref{fig:a_hmf_2} compares the HMFs of the \calsim\ runs that include both baryon physics and massive neutrinos (curves) with that expected if the feedback and free-streaming are treated separately (crosses).  Specifically, we characterise the suppression due to AGN feedback alone as the ratio of the HMF of the \calsim\ hydro run with massless neutrinos (\nua) to that of the \calsim\ \dmonly\ run with massless neutrinos (\nua~DM), and we characterise the suppression due to neutrino free-streaming alone as the ratios of the various \calsim\ \dmonly\ runs with massive neutrinos (\nua~DM, \nub~DM, \nuc~DM, \nud~DM, and \nue~DM) to the \calsim\ \dmonly\ run with massless neutrinos (\nua~DM).  We then multiply these separate suppression factors to obtain the combined suppression, such that the multiplicative prediction for the HMF is given by:
\begin{equation}
 \Phi_{NU\ X}^{Mult} = \Phi_{NU\ 0\ DM} \cdot \left(\frac{\Phi_{NU\ X\ DM}}{\Phi_{NU\ 0\ DM}}\right) \cdot \left(\frac{\Phi_{NU\ 0}}{\Phi_{NU\ 0\ DM}}\right)
 \label{eq:hmf_mult}
\end{equation}

\noindent where NU\ X\ DM is the chosen collisionless run with massive neutrinos.

\begin{figure}
 \includegraphics[width=0.995\columnwidth]{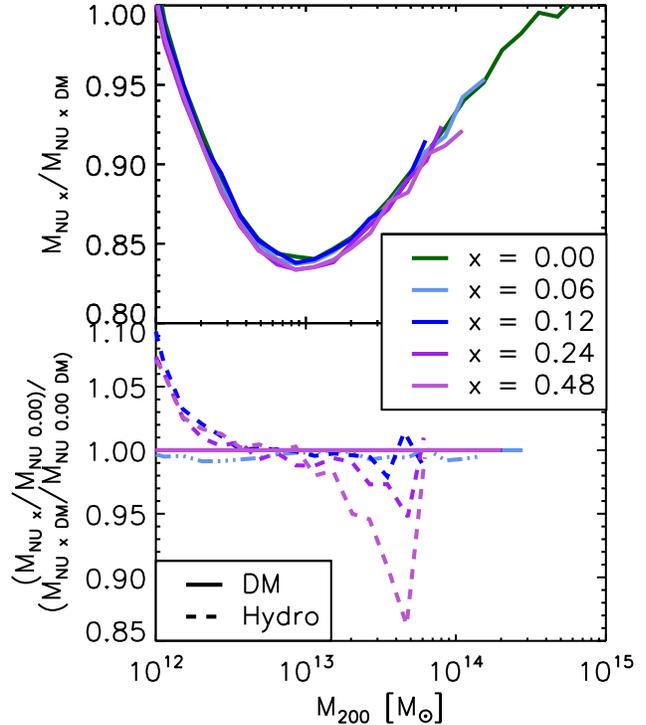}
 \caption{Two tests of the separability of the effects of neutrino free-streaming and baryon physics on halo mass.  {\it Top:} The effect on the halo mass due to baryon physics at different fixed values of the summed neutrino mass for a matched set of haloes. Colours correspond to the different values for the summed neutrino mass. The effect of baryon physics on the halo mass is independent of the choice of summed neutrino mass to approximately 1\% accuracy.  {\it Bottom:} The effect on the halo mass due to neutrino free-streaming for different physics models, normalised to the collisionless case. Solid and dashed lines correspond to the dark matter-only results and those of the  \calsim~feedback model respectively. The effect of neutrino free-streaming is independent of the implemented baryon physics. 
 \label{fig:mass_change}}
\end{figure}

 The bottom panel of Fig.~\ref{fig:a_hmf_2} shows the ratio of the HMF of the self-consistent simulations i.e., with both baryons and neutrinos present together) to that predicted by the treating the baryons and neutrinos separately (i.e., we take the ratio of the lines and crosses in the top panel of Fig.~\ref{fig:a_hmf_2}).  As can be seen, multiplying the separate effects of baryon physics and neutrino free-streaming reproduces their combined effect obtained when both are included simultaneously remarkably well.  The self-consistent HMFs are reproduced to a few percent accuracy by combining the separate effects of neutrinos and baryons in a multiplicative fashion\footnote{We note that we have also experimented with combining the separate effects of neutrinos in {\it additive} fashion, by adding the mass loss due to baryons alone to that from neutrino free-streaming alone and comparing the resulting halo mass function with that derived from the self-consistent simulations with both effects present simultaneously.  We find, however, that this generally results in a poorer reproduction of the HMF predicted by the self-consistent simulations, whereas the multiplicative treatment works very well over all mass ranges.} over the full range of halo masses, summed neutrino masses, and redshifts that we consider.

It is perhaps somewhat surprising that the effects of baryon physics and neutrinos can be treated independently in this way.  For example, an implication of eqn.\ (1) is that a halo whose mass has been reduced (relative to a dark matter-only sim) by neutrino free-streaming is not any more susceptible to gas expulsion by AGN feedback than a halo of the same mass in a massless neutrino case. In other words, the effects of baryon physics or neutrino free-streaming in a simulation with both present are almost of the same magnitude as when one of these processes is omitted.

\begin{figure*}
 \includegraphics[width=0.80\textwidth]{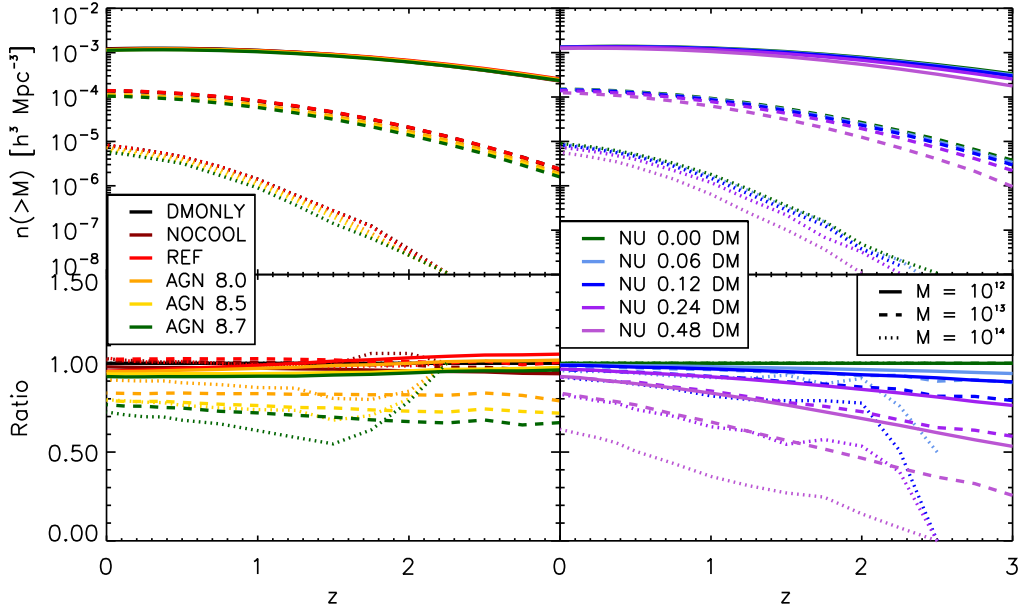}
 \caption{\label{fig:a_CC_1}
 Evolution of the comoving halo space densities above different mass thresholds [i.e., $n(M_{200,{\rm crit}}>M_{\rm threshold},z)$] for the different baryon physics runs in the absence of neutrino physics using cosmo-OWLS (left) and in the absence of baryon physics using the different collisionless massive neutrino runs from \calsim\ (right).  Solid, dashed, and dotted curves correspond to threshold masses of $10^{12}~{\rm M_{\odot}}$, $10^{13}~{\rm M_{\odot}}$, and $10^{14}~{\rm M_{\odot}}$, respectively.  The bottom left panel shows the halo space densities for the baryon physics models normalised to the \dmonly\ case, while the bottom right panel shows the massive neutrino models normalised to the dark matter-only massless neutrino (\nua~DM) case.  The introduction of AGN feedback results in a suppression of the halo space density that is nearly independent of redshift, while the suppression above a fixed mass threshold due to \nufs\ increases strongly with increasing redshift, particularly for models with high values of the summed neutrino mass.  }
\end{figure*}

We explore the separability of baryon physics and neutrino free-streaming further in Fig.~\ref{fig:mass_change}.  In the top panel we show the effect of baryon physics on the  halo mass at different fixed values of the summed neutrino mass.  That is, for a fixed value of the summed neutrino mass, we compare (take the ratio of) the masses of a matched set of haloes in the hydrodynamical and dissipationless simulations.  We plot the median ratio in bins of halo mass.

To an accuracy of approximately 1 percent, the effect on the median halo mass due to baryon physics is independent of the choice of summed neutrino mass.

In the bottom panel of  Fig.~\ref{fig:mass_change} we show the effect on the halo mass due to neutrino free-streaming for two different physics models: the dark matter-only case and the \calsim~calibrated feedback model.  Here we again see that the effects of neutrino free-streaming are nearly independent of the included baryon physics.  Thus, the level of accuracy with which the simple separability assumption reproduces the self-consistent neutrinos+baryon physics simulations in Fig.~\ref{fig:a_hmf_2} is no coincidence, it reflects the fact that these processes truly are approximately independent of one another.

\subsection{Cluster Counts}
 In Fig.~\ref{fig:a_CC_1} we show the effects of baryon physics and \nufs\ on the halo space density for haloes with masses  exceeding different threshold values of $10^{12}~{\rm M_{\odot}}$, $10^{13}~{\rm M_{\odot}}$, and $10^{14}~{\rm M_{\odot}}$.  At a given redshift, the space density is computed by simply integrating the HMF above a given mass threshold\footnote{Note that because of the steepness of the HMF, the total halo space density is dominated by haloes with masses near the chosen threshold value.}.  The halo space density, or `number count', is more closely linked to what is typically measured observationally, as many surveys do not have a sufficiently large volume to robustly measure the HMF, particularly at high masses.

\begin{figure*}
 \includegraphics[width=0.80\textwidth]{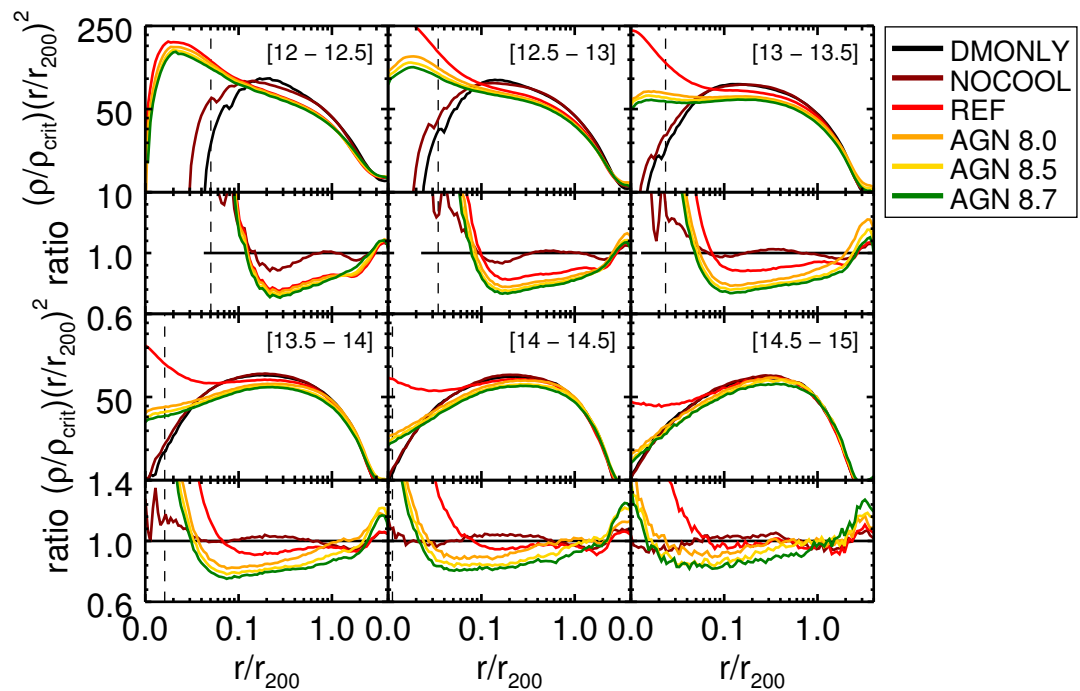}
 \caption{Median  radial total mass density profiles in 0.5 dex mass bins for different baryon physics models in the absence of neutrino physics at fixed cosmology.  Individual panels correspond to mass bins in the dark matter-only, massless neutrino simulation with the stated ranges of $\log(M_{200,{\rm crit}}^{\dmonly}/{\rm M_\odot})$. Haloes in other simulations were binned using the mass of their matched DM only, massless neutrino equivalent and $r_{200}$ corresponds to that from the DM only, massless neutrino run. Line colours correspond to runs with different subgrid prescriptions for baryon physics as in \textcolor{black}{F}igs\textcolor{black}{.} \ref{fig:a_hmf_1} and \ref{fig:a_CC_1}. The vertical dashed line marks the location of three times the gravitational softening length from the halo center. The inclusion of baryonic cooling results in much higher central densities while leaving the outskirts largely untouched compared to the NOCOOL case. The introduction of AGN heating redistributes material from the central regions ($r/r_{200}<0.1$) to the outskirts ($r/r_{200} > 0.5$). This effect is greatest at low halo masses.
 \label{fig:rho_AGN}}
\end{figure*}

\begin{figure*}
 \includegraphics[width=0.80\textwidth]{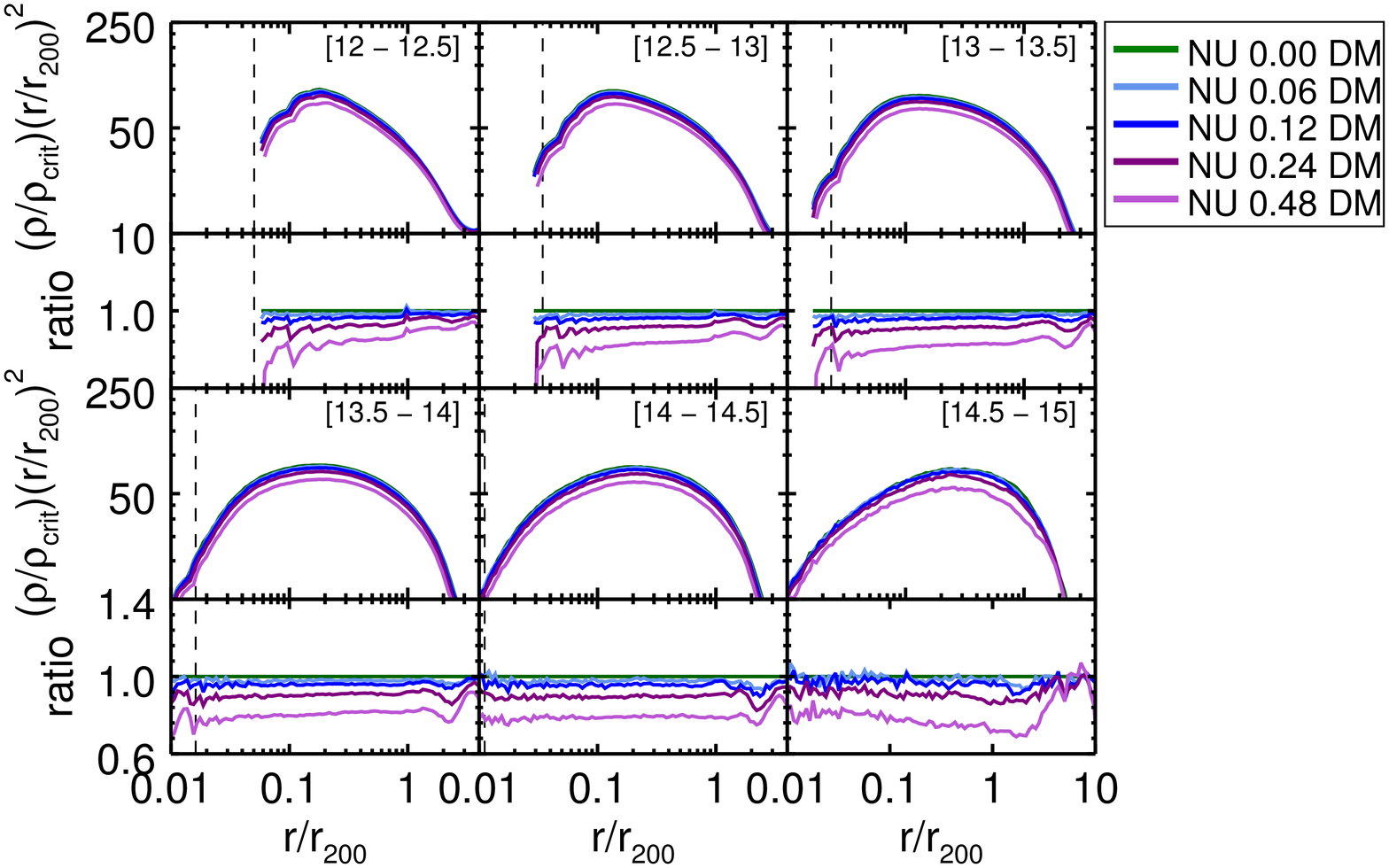}
 \caption{Median  radial total mass density profiles in 0.5 dex mass bins for different neutrino physics models in the absence of baryonic physics at fixed cosmology.  As in Fig. \ref{fig:rho_AGN}, the  panels correspond to different mass bins with the stated ranges in $\log(M_{200,{\rm crit}}^{\dmonly}/{\rm M_\odot})$ and haloes are binned on the mass of the matched DM-only, massless neutrino halo. Different neutrino masses are denoted by colour as in Figs \ref{fig:a_hmf_1} and \ref{fig:a_CC_1}. The vertical dashed line marks the location of three times the gravitational softening length from the halo center.   Neutrino free-streaming lowers the amplitude of the mass density profiles while approximately preserving their NFW-like shape (within the virial radius).
 \label{fig:rho_nu}}
\end{figure*}

The top panels of Fig.~\ref{fig:a_CC_1} demonstrate that the evolution of the halo space density is sensitive to baryon physics and the presence of massive neutrinos, although the dependencies on halo mass and redshift are clearly stronger.  In the bottom panels of Fig.~\ref{fig:a_CC_1}, we effectively remove the halo mass dependence by showing the ratio of the halo space density with respect to that predicted by the \dmonly\ case (left) or with respect to the \nua~DM case (right) for the different mass thresholds.  The bottom left panel of Fig.~\ref{fig:a_CC_1} shows that AGN feedback reduces the abundance of haloes of fixed mass (as shown previously, e.g., \citealt{Cusworth2014,Velliscig2014}).  We find that the suppression does not evolve significantly with redshift.  By contrast, the abundance of haloes above a fixed mass threshold becomes increasingly suppressed at high redshift by neutrino free-streaming, particularly for high mass thresholds and high summed neutrino masses (bottom right panel of Fig.~\ref{fig:a_CC_1}).  We can understand the latter result  by recognizing that by considering haloes above a fixed mass threshold, we are considering increasingly rare systems with lower initial overdensities when moving to higher redshift (e.g., a $10^{13}~{\rm M}_{\odot}$ at $z=2$ will correspond to a massive cluster today).

As was the case for the HMF, we find that the combined effects of baryon physics and neutrino free-streaming on the integrated halo space density can be recovered to a few percent accuracy by treating the baryon physics and neutrino effects separately (i.e., multiplicatively), but for brevity we do not show this here.

\section{Halo structure}

Having explored the separate and combined effects of feedback and massive neutrinos on the overall abundance of haloes, we now examine their effects on the internal structure of haloes.  In particular, we examine the spherically-averaged density profiles in bins of halo mass and the halo mass$-$concentration relation.

\subsection{Total mass density profiles} \label{sec:rho}

In Figs.~\ref{fig:rho_AGN} and \ref{fig:rho_nu} we plot the median total mass density (including stars, gas, and dark matter) profiles in bins of halo mass.  Each panel corresponds to a different halo mass range (each 0.5 dex in width), ranging from $\log(M_{200,{\rm crit}}/{\rm M}_\odot) =$ $13$ to $15.5$ (top left to bottom right).  Fig.~\ref{fig:rho_AGN} shows the effects of baryon physics in the absence of massive neutrinos on the total mass density profiles, while Fig.~\ref{fig:rho_nu} shows the effects of neutrino free-streaming in the absence of baryon physics.  To reduce the dynamic range of the plots, we have scaled the mass density by $r^2$ (i.e., so that an isothermal distribution would correspond to a horizontal line).  Note that the subpanels show the profiles normalised to that predicted by the \dmonly\ case (Fig.~\ref{fig:rho_AGN}) or the dark matter-only case with massless neutrinos (i.e., \nua~DM; Fig.~\ref{fig:rho_nu}).

Because baryon physics and \nufs\ alter the masses of haloes, a selection based on the masses extracted directly from each of the simulations will in general result in somewhat different samples of haloes from the different simulations in a given mass bin.  This is not ideal for our immediate purpose, since our aim is to isolate the physical effects of feedback and neutrinos on a given set of haloes.  We therefore first select haloes from the \dmonly\ (Fig.~\ref{fig:rho_AGN}) and \nua~DM (Fig.~\ref{fig:rho_nu}) runs and then identify the corresponding haloes in the baryon and massive neutrino runs using the unique particle IDs for the dark matter particles as discussed in Section \ref{sec:HMF}.  Therefore, the various panels in Figs.~\ref{fig:rho_AGN} and \ref{fig:rho_nu} {\it correspond to bins of total mass in the dark matter-only, massless neutrino simulations for a matched set of haloes}. Similarly, the values of $r_{200}$ employed in the construction of the profiles are those of the matching DM-only, massless neutrino case halo.

Fig.~\ref{fig:rho_AGN} shows that the inclusion of baryonic physics can significantly alter the radial total density profile away from the standard NFW shape. In the absence of radiative cooling and AGN feedback, the baryons closely trace the dark matter resulting in minimal alteration to the profile (e.g., \citealt{Lin2006}). However, the activation of radiative cooling, star formation and stellar feedback causes much higher central densities with a corresponding reduction in the density between 0.08 and 1 $r_{200}$ of $\approx10\%$, as seen by comparing the \refsim\ and \nocool\ cases. AGN heating somewhat counteracts this effect, reducing central densities while redistributing material to the outer regions of the halo. While the density profiles of all three AGN models examined here are similar between 0.08 and 1 $r_{200}$, higher values for AGN heating result in higher densities beyond $r_{200}$ and lower densities in the central regions of the halo.  The redistribution of material causes the scale radius to increase relative to the \dmonly\ case.  This makes intuitive sense as the more energetic (albeit comparatively infrequent) outbursts of AGN with higher heating temperatures will eject more mass from the progenitors of the halo and cause a greater degree of expansion of the dark matter.  The effects of baryonic physics become less important in higher mass bins, due to the deeper potential wells of these systems.

Fig.~\ref{fig:rho_nu} shows that, to a first approximation, the impact of neutrino free-streaming alone (i.e., with no baryons present) is to lower the overall amplitude of the mass density profiles within $r_{200}$ while approximately preserving the NFW-like shape.  In effect, the free-streaming of massive neutrinos acts primarily to reduce the mass of a given halo. Beyond $\sim r_{200}$, however, there is also a change in shape, as is evident from the `oscillatory' feature in the subpanels that show the ratio of the profiles with respect to that of the massless neutrino case.  Physically, we interpret this feature as being due to the less evolved state of collapse of clusters in the simulations with massive neutrinos.  In the language of clustering, the scale that marks the transition from the `1 halo' term (i.e., the profile of the central halo) to the `2 halo' term (the clustering of other nearby systems), as well as its amplitude, is altered by neutrino free-streaming.  We plan to explore the use of this feature as a constraint on the summed neutrino mass in a future study.

\begin{figure*}
 \includegraphics[width=0.80\textwidth]{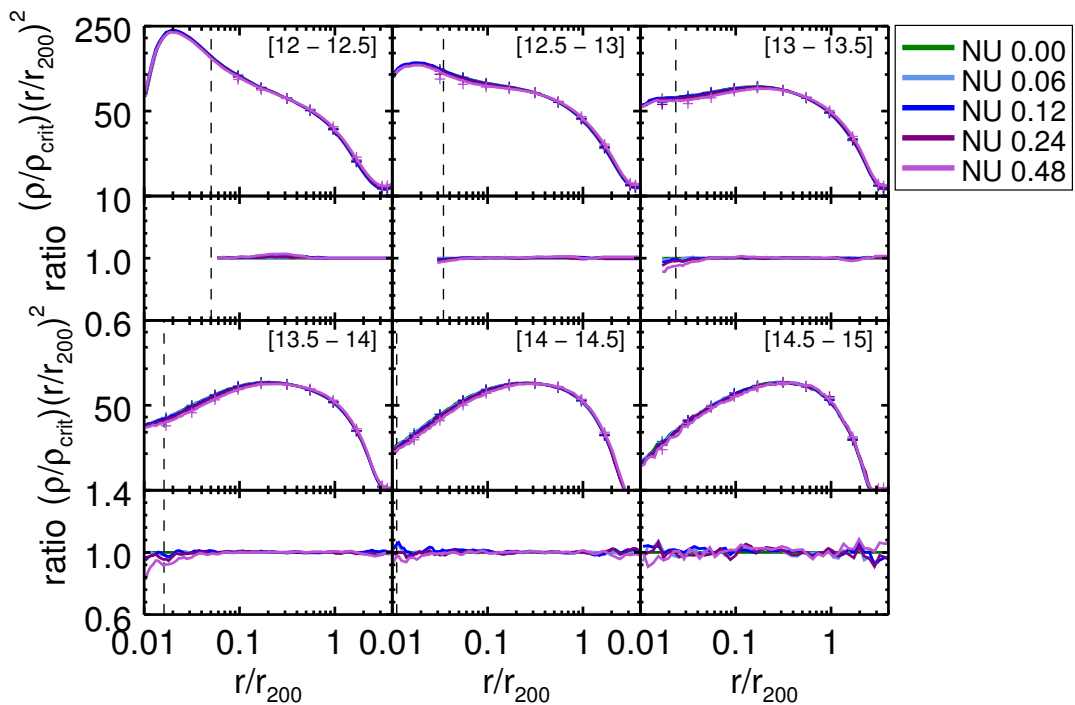}
 \caption{Comparison of the median  radial total mass density profiles of haloes arising when simulating baryonic feedback and neutrino free-streaming simultaneously (curves) and with that calculated by multiplying their separate effects (crosses). Haloes are binned using their self-consistent masses from each of the simulations, with the mass ranges for each bin stated in units of $\log(M_{200,{\rm crit}}/{\rm M}_\odot)$. Line colours correspond to runs with different neutrino masses as in Figs \ref{fig:a_hmf_2} and \ref{fig:mass_change}, and the vertical dashed line shows the location of three times the gravitational softening length from the halo center. The two cases agree to within a few percent at $r>0.05r_{200}$, with the exception of the highest mass bin where we have relatively poor statistics. 
 \label{fig:rho_Mult}}
\end{figure*}

As in Section 3, we investigate to what degree the two processes may be treated independently. In Fig.~\ref{fig:rho_Mult} we compare the results of simulation runs combining baryon physics and neutrino free-streaming (curves) with those obtained by multiplying together the strengths of the two effects in isolation, this time for the radial density profiles (crosses).  Our formalism for this is identical to that shown in eq.~(\ref{eq:hmf_mult}), with the HMF exchanged for the radial density profile: 
\begin{multline}
  \rho(r)_{NU\ X}^{Mult} = \rho(r)_{NU\ 0\ DM} \cdot \\ \left(\frac{\rho(r)_{NU\ X\ DM}}{\rho(r)_{NU\ 0\ DM}}\right) \cdot \left(\frac{\rho(r)_{NU\ 0}}{\rho(r)_{NU\ 0\ DM}}\right)
 \label{eq:hmf_rho}
\end{multline}

\noindent As can be seen, the combined effects are reproduced to an accuracy of a few percent in all but the very central regions of the halo ($r<0.05 r_{200}$), with the exception of the highest \textcolor{black}{m}ass bin where we have comparatively poor statistics.

It is important to note here that in Fig.~\ref{fig:rho_Mult} we have reverted back to an unmatched set of haloes.  That is, we have used the self-consistent masses from each of the simulations for this test.

\subsection{Mass$-$concentration relation}

The internal structure of CDM haloes in cosmological simulations is known to depend on their formation history, in that systems that collapsed earlier tend to have higher present-day concentrations on average than those that collapsed later on (e.g., \citealt{Wechsler2002}).  This sensitivity is linked to the evolution of the (background) density of matter in the Universe, such that systems that collapsed earlier on had to have a higher physical density (in an absolute sense) to be overdense with respect to the background density, which was higher at earlier times.  In CDM models, low-mass haloes typically collapse before high-mass haloes and, when combined with the evolution of the background density, this gives rise to the expectation that low-mass systems ought to be more concentrated than high-mass haloes, a result which is borne out in high resolution cosmological simulations (e.g., \citealt{Bullock2001,Eke2001,Neto2007}).

As we have seen in Section \ref{sec:rho}, non-gravitational processes (e.g., feedback) and neutrino free-streaming also alter the internal structure of collapsed haloes, and therefore ought to modify the mass$-$concentration relation.  Here we examine the separate and combined effects that these processes have on this relation.

\begin{figure*}
 \includegraphics[width=0.80\textwidth]{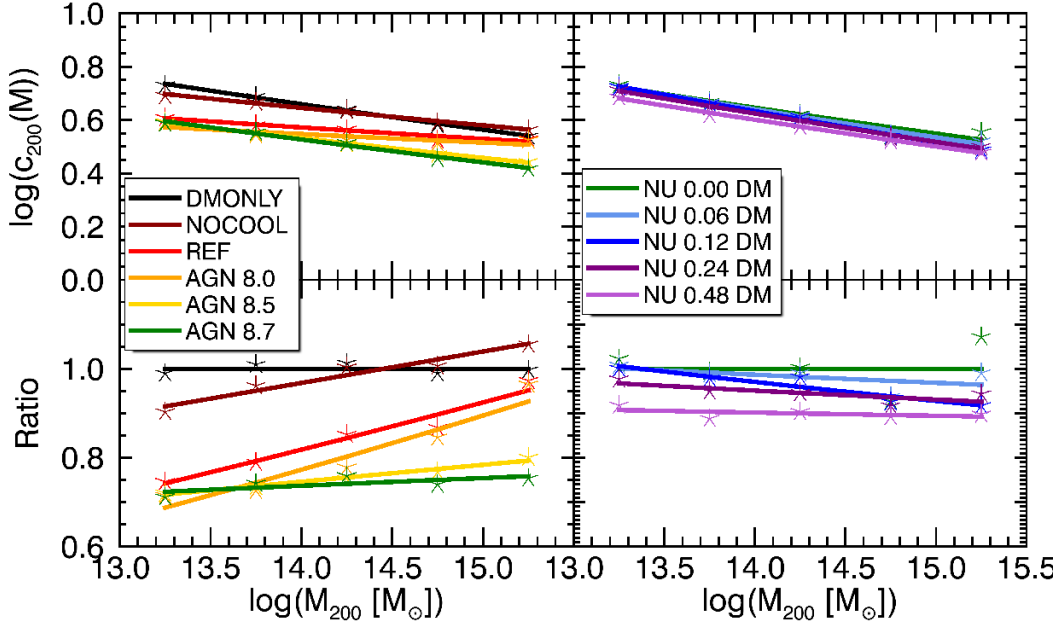}
 \caption{\label{fig:a_MC_1}
 Best fit total mass \mc\ relations for different baryon physics models in the absence of neutrino physics in the \wmapa\ cosmology (left) and for different \mnu\ values in the absence of baryonic physics in \wmapb\ cosmology (right) at $z=0$. Halo masses are those from each of the individual simulations (i.e., not the matched DM-only masses). Stars mark the locations of the mean concentration value in each 0.5 dex mass bin, solid lines display the best-fit power laws to these means with the functional form of equation \ref{eq:Neto}.  The upper panel displays these in log(c) - log(M) space, while the lower panel displays the same data normalised to the best fit for the the relevant DM only model. Increasing \mnu\ results primarily in a reduction of the amplitude of \mc\ with respect to the \dmonly~\nuadm\ model with minimal alteration to the gradient. Conversely, \fdbk\ alters both the amplitude and the gradient.  }
\end{figure*}

As is customary, we define the concentration parameter, $c_{\Delta}$, as the ratio of the radius enclosing an overdensity of $\Delta$ times the critical density, $r_{\Delta}$ to the NFW scale radius, $r_s$ (i.e., $c_{\Delta} \equiv r_{\Delta}/r_s$).  We derive two estimates of the scale radius (and therefore the concentration), by fitting NFW profiles to the total and dark matter mass density profiles, respectively.  Below we present results for the case of $\Delta=200$.  When deriving estimates of the scale radius, we fit the NFW profile over the radial range $0.1 \le r/r_{200} \le 1.0$ for halo masses of $\log(M_{200,{\rm crit}}/{\rm M_{\odot}}) \ge 13$, noting that by adopting a minimum radius of $0.1\ r_{200}$ we largely avoid the region dominated by stars that is typically not well fit by the NFW form for these haloes. We exclude haloes below this mass as the star dominated region approaches the scale radius (see Fig.~\ref{fig:rho_AGN}).\footnote{Note that $r_s  > 0.1\ {\rm r_{200}}$ for the halo mass ranges under consideration\textcolor{black}{.}}. To give approximate equal weighting to the different radial bins over the range that we consider, we actually fit to the quantity $\rho \ r^2$, as done in several previous studies (e.g., \citealt{Neto2007}).  We derive concentration estimates for each individual halo satisfying $M_{200,{\rm crit}} > 10^{13} {\rm M_\odot}$ in all of the simulations.

The resulting \conc\ values are binned into equally-spaced logarithmic mass bins (0.5 dex width) between $13.0$ and $15.0$ in $\log{(M_{200,{\rm crit}}/{\rm M_\odot})}$.  In each bin we calculate the mean concentration value, \mconc, the standard deviation ($\sigma_{ln(c_{200})}$) of the intrinsic scatter around \mconc, and the mean halo mass, $\langle M_{200,{\rm crit}} \rangle$.  As the scatter in \conc\ is approximately log normal, \mconc\ and $\sigma_{ln(c_{200})}$ were computed by fitting a Gaussian distribution to the histogram of the \conc\ values in $100$ equally-spaced logarithmic bins spanning $3$ dex centred on an estimate for the mean value.  

Previous studies found that the distribution of mass and concentration values for \dmonly~haloes in N-body simulations at $z~=~0$ was well fitted by a power law of the form
\begin{equation}
 c_{\Delta}(M_{\Delta}) = A\cdot \left(\frac{M_{\Delta}}{M_{Fiducial}}\right)^{B}\,\, .
 \label{eq:Neto}
\end{equation}

\noindent \citet{Gao2008} demonstrated that a power law of this form continued to be a good fit to samples of haloes in N-body simulations out to redshifts of 2, although the value of the parameter $A$ varied as a function of $z$.  We follow \citet{Duffy2008} and parametrise this redshift dependence by expanding Eqn.~\ref{eq:Neto} to the form 
\begin{equation}
 c_{\Delta}(M_{\Delta}) = A\cdot \left(\frac{M_{\Delta}}{M_{Fiducial}}\right)^{B}\cdot(1+z)^{C} \,\, .
 \label{eq:Duffy}
\end{equation}
Note that at fixed redshift, the $A$ and $C$ parameters in Eqn.~\ref{eq:Duffy} are degenerate.  Therefore, when we present the results of our analysis below at $z=0$ we present fits to Eqn.~\ref{eq:Neto}.  However, we include the results of fitting Eqn.~\ref{eq:Duffy} over the redshift range $0 \leq z \leq 2$ in Table \ref{tab:DuffyABC} in the Appendix.

 Figure \ref{fig:a_MC_1} displays the best-fit $z=0$ total mass$-$concentration relations from equation \ref{eq:Neto} for different baryon physics models in the absence of neutrino physics in the \wmapa\ cosmology (left panel), and for the different \mnu\ values in the absence of baryonic physics in \textcolor{black}{the} \wmapb\ cosmology (right panel), using the self-consistent masses from each simulation (i.e., for an unmatched set of haloes).

\begin{figure*}
 \includegraphics[width=0.80\textwidth]{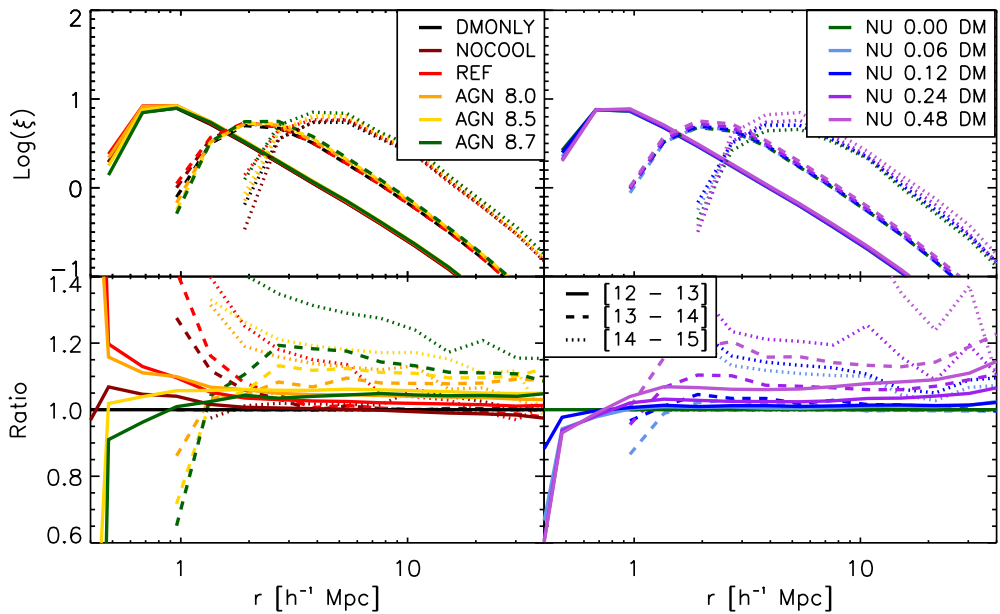}
 \caption{\label{fig:cl_1}
  Real-space 2-point halo autocorrelation functions ($\xi$)\ for the different baryonic physics runs in the absence of neutrino physics from cosmo-OWLS (left) and the different collisionless massive neutrino runs from BAHAMAS (right). Solid, dashed and dotted curves correspond to $\xi$ for haloes in mass bins of $10^{12}-10^{13} {\rm M}_\odot$, $10^{13}-10^{14} {\rm M}_\odot$ and $10^{14}-10^{15} {\rm M}_\odot$ in $M_{200,{\rm crit}}$ respectively. The bottom left panel shows $\xi$ for each of the baryonic physics models normalised to the \dmonly~case, while the bottom right panel shows the massive neutrino models normalised to the dark matter-only massless neutrino (\nua\ DM) case. The introduction of AGN feedback results in a $\sim10\%$ increase in the amplitude of the autocorrelation function, with the precise shift depending on the halo mass range and AGN heating temperature under consideration. Neutrino free-streaming also increases the amplitude, with the strength of the effect depending sensitively on the precise value of the summed neutrino mass.}
\end{figure*}

Consistent with previous studies (e.g., \citealt{Duffy2010}), we find that the inclusion of efficient feedback results in a lowering of the amplitude of the mass$-$concentration relation (left panel of Fig.~\ref{fig:a_MC_1}).  There is also a slight shallowing of the relation with respect to the \dmonly\ case, driven by the increasing importance of feedback with decreasing halo mass.  Note that the mass$-$concentration relation is altered in two ways: the profile shapes (and therefore the scale radius) are altered by feedback (Fig.~\ref{fig:rho_AGN}) and the overall halo mass is also affected.  By comparing the results to those for a matched set of haloes (not shown), we find that the main effect is from the increase in the scale radius as opposed to the lowering of the halo mass.

In the right panel of Fig.~\ref{fig:a_MC_1} we see that neutrino free-streaming lowers the amplitude of the mass$-$concentration relation and also has a slight effect on its shape.
As in the case of feedback, this is due both to a (slight) change in the shapes of the profiles (an increase in the scale radius) and to a lowering of the overall halo mass.  By analysing the mass$-$concentration relation for a matched set of haloes (not shown), we deduce that the change in the halo mass is more important than the change in the scale radius for halo masses above $\sim 14.5$ in $\log(M_{200,{\rm crit}}/{\rm M_{\odot}})$, while the reverse is true at lower masses.

In analogy to our exploration of the halo masses, we have examined to what extent the effects of baryon physics and neutrino free-streaming on the mass$-$concentration relation can be treated separately (i.e., does it reproduce the combined effect, when both baryons and massive neutrinos are present).  We find that in a relative sense treating these effects separately reproduces the combined result to a few percent accuracy, as would be expected from the similar success in recovering the density profiles (see Fig. \ref{fig:rho_Mult}). For brevity we do not show this here.

 \section{Halo clustering}

Having quantified the effects of feedback and neutrino free-streaming on the masses and internal structure of haloes, we now proceed to examine their separate and combined effects on the spatial distribution of haloes.  Specifically, we focus here on the clustering of FOF haloes in bins of halo mass, as characterised by the 3D two-point autocorrelation function.  We examine the clustering of matter in general in Section 6.

We compute the autocorrelation function, $\xi$, of FOF groups as the excess probability (with respect to a random distribution) of having another FOF group present at a particular distance; i.e.,
\begin{equation}
 \xi(r) = \frac{DD(r)}{RR(r)} - 1
\end{equation}

\noindent where $DD(r)$ and $RR(r)$ are the `data' and `random' pair counts in radial bins.  We compute $RR$ analytically, assuming the FOF groups are spread homogeneously throughout the simulation volume at the mean density of haloes of the particular mass range under consideration.  We compute $\xi$ in 20 logarithmic radial bins between 0.1 and 100 $h^{-1}$ comoving Mpc.

\begin{figure*}
 \includegraphics[width=0.80\textwidth]{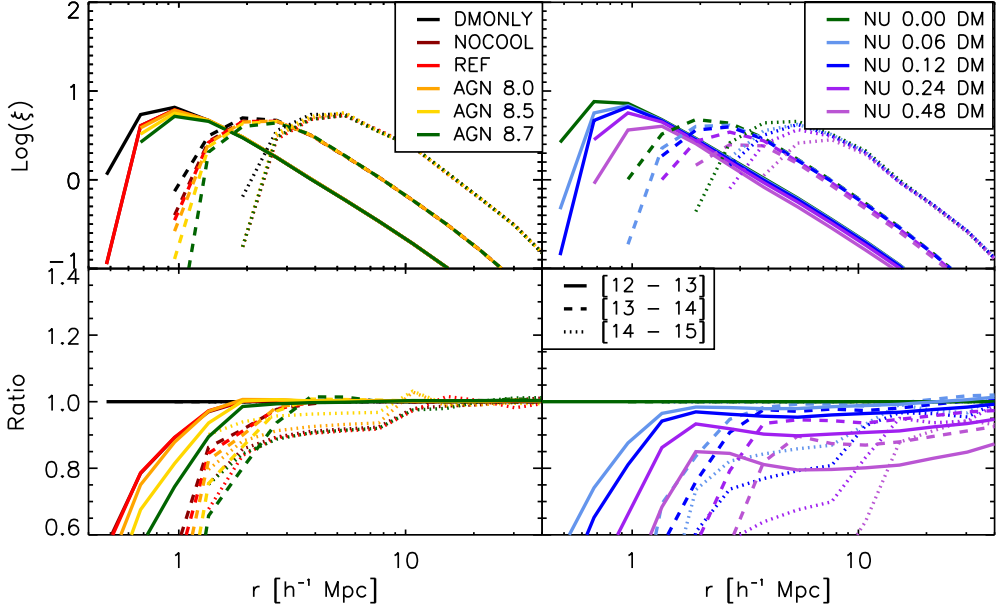}
 \caption{Real-space 2-point halo autocorrelation functions ($\xi$)\ for the different baryonic physics models in the absence of neutrino physics from cosmo-OWLS (left) and the different collisionless massive neutrino runs from BAHAMAS (right). The bottom left panel shows $\xi$ for each of the baryonic physics models normalised to the \dmonly~case, while the bottom right panel shows the massive neutrino models normalised to the dark matter-only massless neutrino (\nua\ DM) case. In contrast to Fig.~\ref{fig:cl_1}, solid, dashed and dotted curves correspond to $\xi$ for \textcolor{black}{matched} haloes in mass bins of $10^{12}-10^{13} {\rm M}_\odot$, $10^{13}-10^{14} {\rm M}_\odot$ and $10^{14}-10^{15} {\rm M}_\odot$ in $M_{200,{\rm crit}}^\dmonly$ respectively, i.e. selecting the same set of haloes in each simulation. As can be clearly seen from the bottom-left panel, the large-scale clustering of a chosen set of haloes is, to a very high level of accuracy, unaffected by baryon physics. Conversely, neutrino free-streaming can suppress the amplitude of the halo autocorrelation function by $\sim10\%$.  \label{fig:cl_1_matched}
}
\end{figure*}

In Fig.~\ref{fig:cl_1} we show the separate effects of baryon physics (left panel) and neutrino free-streaming (right panel) on the autocorrelation in three different halo mass bins.  Consistent with the results of \citet{vanDaalen2014}, we find that AGN feedback {\it increases} the amplitude of the autocorrelation by $\sim10\%$, with the precise shift depending on the halo mass range and the AGN heating temperature.  Neutrino free-streaming has a qualitatively similar effect, with the shift depending sensitively on the adopted mass range and the summed mass of neutrinos, $M_\nu$.

At first sight it is odd that the inclusion of massive neutrinos leads to an increase in the amplitude of the halo clustering signal, given that it is well known that neutrinos suppress the matter power spectrum (e.g., \citealt{Bond1980}, see also Section 6).  The origin of this apparent inconsistency lies in the fact that we are plotting the clustering signal in bins of halo mass in Fig.~\ref{fig:cl_1} and that we are using the self-consistent masses from each of the simulations when placing the FOF groups into halo mass bins.  Since feedback and neutrino free-streaming affect the halo masses (they generally lower them with respect to the \dmonly\ case with massless neutrinos), the clustering signal will be different for different simulations simply because we are considering a different set of systems for each simulation.  Indeed, \citet{vanDaalen2014} have shown that, in the case of massless neutrino simulations, the increased amplitude of the large-scale autocorrelation in hydrodynamical simulations with respect to the \dmonly\ case can be entirely accounted for by the change in halo mass.

We confirm the findings of \citet{vanDaalen2014} in the left panel of Fig.~\ref{fig:cl_1_matched}, where we use our halo matching technique to identify a common set of haloes for the different simulations.  Specifically, we bin haloes by their corresponding masses in the \dmonly\ case.  To a high level of accuracy, we find that on scales $r~\gg~r_{200}$ the clustering is unaffected by baryon physics when considering a common set of haloes (i.e., feedback does not push haloes around).

The situation is different in the case of neutrino free-streaming, however, which we consider in the right panel of Fig.~\ref{fig:cl_1_matched}.  In particular, when we account for the effects of changes in the halo mass, by adopting a common set of haloes, we find that the large-scale clustering signal is now suppressed by $\sim10\%$.  Physically this makes sense, since the free-streaming of the neutrino background acts to delay the growth of fluctuations (i.e., it suppresses the matter power spectrum).  
This result is useful for galaxy surveys that compare with semi-empirical models such as SHAM, or with SA models, which are based on the masses of haloes in \dmonly\ simulations. However, it is important to note that for observational surveys that use directly measured masses, e.g. by combining with galaxy-galaxy lensing, it is Fig.~\ref{fig:cl_1} that is most directly relevant.

\begin{figure}
 \includegraphics[width=0.955\columnwidth]{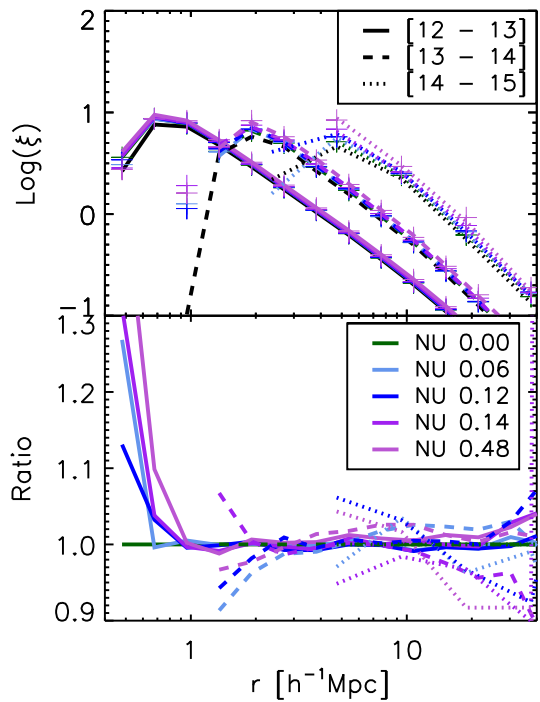}
 \caption{\label{fig:cl_2}
 Comparison of the real space 2-point halo autocorrelation functions ($\xi$)\ arising when simulating baryonic feedback and neutrino free-streaming simultaneously (curves) and those calculated by multiplying the separate effects of baryonic feedback in the absence of neutrinos and the effects of neutrino free-streaming in the absence of baryon physics (crosses). Solid, dashed and dotted lines correspond to  $\xi$ for haloes in mass bins of $10^{12}-10^{13} {\rm M_\odot}$, $10^{13}-10^{14} {\rm M_\odot}$ and $10^{14}-10^{15} {\rm M_\odot}$ in $M_{200,{\rm crit}}$ respectively. The bottom panel shows each of the multiplicative models normalised by the corresponding combined case. The multiplicative treatment recovers the combined result with a few percent accuracy for $r > 1 {\rm h^{-1}Mpc}$ independent of the chosen summed neutrino mass in all but the highest mass bin.}
\end{figure}

In Fig.~\ref{fig:cl_2} we test how well treating the effects of baryon physics and neutrino free-streaming separately (i.e., multiplicatively) reproduces their combined effects on the clustering of massive haloes.  For this test, we use the self-consistent halo masses from each simulation, rather than identifying a common set of haloes and binning using masses from the massless \dmonly\ run.  The top panel of Fig.~\ref{fig:cl_2} compares the clustering signal measured directly from the hydrodynamics+neutrino simulations (curves) to that predicted by treating these two processes separately (crosses).  So, for example, the prediction for the \nud~case would be to multiply the clustering signal of the massless \dmonly\ run, \nua~DM, by the ratio of the hydrodynamics to \dmonly\ case with massless neutrinos (i.e., \nua/\nua~DM) and by the ratio of the massive to massless neutrino cases in the absence of baryon physics (i.e., \nud~DM/\nua~DM).  The bottom panel of Fig.~\ref{fig:cl_2} shows the ratio of the prediction arising from the multiplicative approach to the self-consistent calculation.

We find that for $r > 1 {\rm h^{-1} Mpc}$ and $12 \le \log(M_{200,{\rm crit}}/{\rm M_{\odot}}) \le 14$, the combined effects of neutrino free-streaming and baryons physics can be reproduced extremely well (1-2\% accuracy), by considering these effects separately, independent\textcolor{black}{ly} of the choice of summed neutrino mass, $M_\nu$. This agreement worsens slightly in our highest mass bin, where the two cases deviate by \textcolor{black}{$\approx 10\%$} at large radii for the highest summed neutrino mass.

\section{Matter clustering}

As a final LSS diagnostic, we now consider the effects of baryon physics and massive neutrinos on the total matter power spectrum.  We compute the matter power spectra using the GenPK code\footnote{https://github.com/sbird/GenPK}.

\begin{figure*}
 \includegraphics[width=0.80\textwidth]{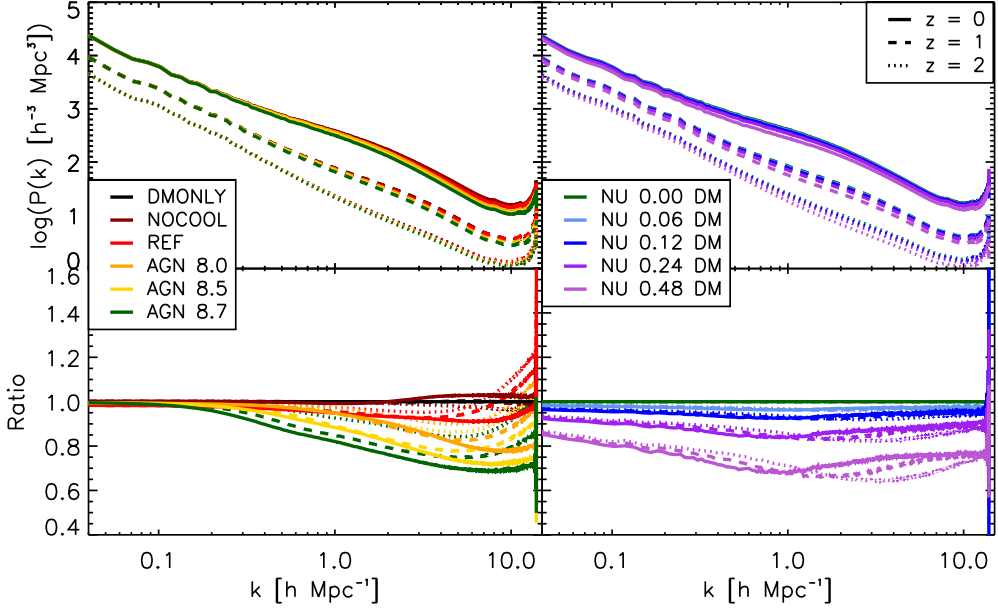}
 \caption{\label{fig:Pk_1}
 Matter power spectra for different baryon physics models in the absence of neutrino physics in the \wmapa\ cosmology (left) and for different \mnu\ values in the absence of baryonic physics in the \wmapb\ cosmology (right) at $z=0$. As in Fig.~\ref{fig:a_hmf_1}, colours denote the various runs (see the legend and Table \ref{tab:sims}), while the different linestyles denote different redshifts. The bottom left panel shows matter power spectra for the cosmo-OWLS runs normalised to the \dmonly\ case, whereas in the bottom right panel the collisionless BAHAMAS runs have been normalised by the massless neutrino case. Baryonic feedback suppresses the matter power spectrum by 10-20\%\ on small scales ($k \ga 1$ h/Mpc). In contrast, the suppression due to neutrino free-streaming depends strongly on the choice of summed neutrino mass and has an effect over a much wider range of scales. The suppression due to baryonic feedback grows by a factor of $\sim 2$ between $z=2$ and $z=0$, whereas the level of suppression resulting from neutrino free-streaming is only weakly dependent on redshift. }  
\end{figure*}

In Fig.~\ref{fig:Pk_1} we show the separate effects of baryon physics (left panel) and neutrino free-streaming (right panel) on the matter power spectrum at three different redshifts.  Consistent with the previous findings of \citet{vanDaalen2011}, we find that AGN feedback suppressed the matter power spectrum on small scales ($k \ga 1$ h/Mpc), at levels of up to 10-20\%.  Neutrino free-streaming also suppresses the matter power spectrum, but over a wider range of scales (up to the free-streaming scale $\sim 100 \ {\rm Mpc}$, \citealt{Bird2013}). While the suppression due to neutrino free-streaming is largely insensitive to the redshift, that due to baryonic feedback grows by a factor of $\sim2$ between $z=2$ and $z=0$ 

In Fig.~\ref{fig:Pk_2} we show the combined effects of baryon physics and neutrino free-streaming (curves) and compare this with the predicted power spectra when these effects are treated separately and then multiplied (crosses).  The predictions reproduce the power spectra from the self-consistent simulations to typically better than 2\% accuracy over the full range of redshifts and summed neutrino masses we have considered for wavenumbers of $k \la 10$ h/Mpc.    This result provides some reassurance for existing studies that have treated these processes independently (e.g., \citealt{HarnoisDeraps2015}).

  \section{Summary and Discussion}                 \label{sec:summary}

We have used the cosmo-OWLS and \calsim~suites of cosmological hydrodynamical simulations to explore the separate and combined effects of baryon physics (particularly AGN feedback) and neutrino free-streaming on different aspects of large-scale structure (LSS), including the halo mass function and halo number counts, the spherically-averaged density profiles and mass$-$concentration relation, and the clustering (autocorrelations) of haloes and matter.

\begin{figure}
 \includegraphics[width=0.955\columnwidth]{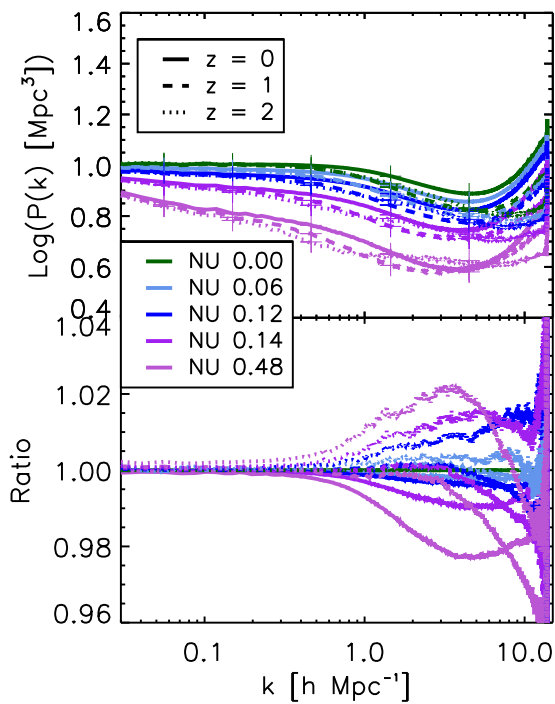}
 \caption{\label{fig:Pk_2}
 Comparison of the matter power spectra (P(k)) arising when simulating baryonic feedback and neutrino free-streaming simultaneously (curves) and those calculated by multiplying the separate effects of baryonic feedback in the absence of neutrinos and the effects of neutrino free-streaming in the absence of baryon physics (crosses). Solid, dashed and dotted lines correspond to P(k) at redshifts of 0, 1 and 2 respectively. The bottom panel shows each of the multiplicative models normalised by the corresponding combined case. As in Fig.~\ref{fig:cl_2}, the multiplicative treatment recovers the combined result to better than 3\% accuracy independent of the chosen summed neutrino mass for $0.04 \leq k {\rm [h/Mpc]} \leq 10 $.}
\end{figure}

From this investigation we conclude the following:
\begin{itemize}
 \item{AGN feedback can suppress the halo mass function by $\textcolor{black}{\approx}20 - 30\%$ relative to the \dmonly\ case on the scale of galaxy groups and clusters, a result which is largely insensitive to redshift (Fig.~\ref{fig:a_hmf_1}, left panels), as also found by \citet{Velliscig2014}.  Neutrino free-streaming preferentially suppresses the high-mass (cluster) end of the HMF, with a strong dependence on redshift and the choice of summed neutrino mass (Fig.~\ref{fig:a_hmf_1}, right panels).}  
 \item{In terms of mass density profiles, the inclusion of baryonic physics, and in particular radiative cooling and AGN heating, produces higher central (due to cooling) and peripheral densities (due to gas ejection), with lower densities at intermediate radii (due to gas ejection), relative to the \dmonly~case.  The gas expulsion leads to an expansion} \textcolor{black}{of} the dark matter, such that the NFW scale radius increases (Fig.~\ref{fig:rho_AGN}). To a first approximation, the free-streaming of massive neutrinos reduces the amplitude of the mass density profiles while approximately preserving their shape within the virial radius (Fig.~\ref{fig:rho_nu}).  However, there is a change in the shape of the profile just beyond the virial radius, such that the radius that marks the transition from the `1 halo' to the `2 halo' term decreases with increasing summed neutrino mass.
 \item{Free-streaming of massive neutrinos results in a reduction in the amplitude of the mass$-$concentration relation by $\sim10\%$ (depending on the summed neutrino mass) with only minimal alteration to its slope (Fig.~\ref{fig:a_MC_1}, right panels).  This is due both to a lowering of the overall halo mass and a slight increase of the scale radius.  By contrast, AGN feedback alters both the amplitude and the slope of the mass$-$concentration relation (Fig.~\ref{fig:a_MC_1}, left panels), as also found by \citet{Duffy2010}.  The amplitude shift here is due mainly to an increase in the scale radius, driven by the expansion of the dark matter halo due to gas expulsion from feedback.  The change in slope reflects the increased importance of feedback for groups relative to clusters.}
 \item{In bins of halo mass, both AGN feedback and neutrino free-streaming result in an apparent enhancement of the amplitude of the 2-point halo correlation function on large scales ($r \gg r_{200}$), by $\sim10\%$ with respect to the \dmonly~case with massless neutrinos at $z=0$ (Fig.~\ref{fig:cl_1}).  In the case of simulations with baryons and massless neutrinos, this is due entirely to the effect on the halo mass (so that the mass bins contain different systems in different simulations) rather than a true alteration of the spatial distribution of haloes (Fig.~\ref{fig:cl_1_matched}, left panels), consistent with the findings of \citet{vanDaalen2014}.  In the case of simulations with massive neutrinos, when we account for the change in halo mass we find that the large-scale clustering of haloes is actually suppressed relative to a massless neutrino case (Fig.~\ref{fig:cl_1_matched}, right panels), as expected.}
 \item{On small scales ($k \ga 1$ h/Mpc) the matter power spectrum can be suppressed by AGN feedback by up to 10-20\% at $z=0$, consistent with the previous findings of \citet{vanDaalen2011}. This factor increases by a factor of $\textcolor{black}{\approx} 2$ between $z=2$ and $z=0$ (Fig.~\ref{fig:Pk_1}, left panels).  Neutrino free-streaming also suppresses the matter power spectrum, but over a much wider range of scales (see also \citealt{Semboloni2011}). This suppression is nearly insensitive to redshift but depends strongly on the adopted summed neutrino mass (Fig.~\ref{fig:Pk_1}, right panels).}
\item{We have investigated the extent to which the effects of baryon physics and neutrino free-streaming can be treated independently.  \textcolor{black}{The procedure of m}ultiplying together the magnitudes of the two effects when taken in isolation reproduces their combined effects to typically a few percent accuracy for the halo mass function (Fig.~\ref{fig:a_hmf_2}), the mass density profiles (Fig.~\ref{fig:rho_Mult}), the mass$-$concentration relation, and the clustering of haloes (Fig.~\ref{fig:cl_2}) and matter (Fig.~\ref{fig:Pk_2}) over ranges of $12 \le M_{200,{\rm crit}}/{\rm M}_\odot \le 15$, $12 \le M_{200,{\rm crit}}/{\rm M}_\odot \le 14.5$, $12 \le M_{200,{\rm crit}}/{\rm M}_\odot \le 14$ and $0.04 \le k \ {\rm [h/Mpc]} \le 10$, respectively. Our simulation-based matter power spectrum findings are therefore consistent with those of \citet{Mead2016}, who explored the degeneracies between feedback, massive neutrinos, and modified gravity in the context of a modified `halo model' formalism (see \citealt{Mead2015} for further details).}
\end{itemize}

Our work has demonstrated that both AGN feedback and neutrino free-streaming can have a considerable impact on LSS.  They should therefore both be included in cosmological analyses.  Through the use of self-consistent cosmological simulations we have shown that, to a high degree of accuracy, these processes are separable (i.e., can be treated independently), which should considerably simplify the inclusion of their effects in cosmological studies that adopt, for example, the halo model formalism or the linear matter power spectrum (e.g., from CAMB).

In a future study we will examine the constraints that can be placed on the absolute mass scale of neutrinos from comparisons to current LSS data (McCarthy et al., in prep).

  \section*{Acknowledgements}                  \label{sec:acknowledge}
The authors thank Alex Mead for helpful comments on an earlier version of the paper and Amandine Le Brun for her contributions to the simulations.
BOM is supported by a STFC PhD studentship at Liverpool JMU.  IGM was supported by a STFC Advanced Fellowship.  JS acknowledges support from ERC grant 278594 - GasAroundGalaxies and from the Netherlands Organisation for Scientific Research (NWO) through VICI grant 639.043.409.  SB was supported by NASA through Einstein Postdoctoral Fellowship Award Number PF5-160133.

This work used the DiRAC Data Centric system at Durham University, operated by the Institute for Computational Cosmology on behalf of the STFC DiRAC HPC Facility (www.dirac.ac.uk). This equipment was funded by BIS National E-infrastructure capital grant ST/K00042X/1, STFC capital grants ST/H008519/1 and ST/K00087X/1, STFC DiRAC Operations grant ST/K003267/1 and Durham University. DiRAC is part of the National E-Infrastructure.
 
  \bibliographystyle{mn2e}
  \bibliography{newbib}

\appendix 
 \section{Fits to mass-concentration relations}  

Here we provide the best fit powerlaw parameter values to the mass$-$concentration relations of the various cosmo-OWLS and \calsim~runs.

  \begin{table*}
   \caption{Best\textcolor{black}{-}fit values for the coefficients of eq.~(\ref{eq:Duffy}) for $z = 0-2$ and $M_{Fiducial}=10^{14}{\rm {\rm M_{\odot}}}$, and the $1-\sigma$ log-normal scatter of concentration values around the best fit relation}    
   \begin{tabular}{*{10}{l}}\hline
           &         & \multicolumn{4}{c}{DM}            & \multicolumn{4}{c}{Tot}\\
           &         & A     & B      & C      & scatter & A     & B      & C      & scatter\\
    \hline
           & \nocool & 4.878 & -0.099 & -0.507 & 0.340   & 4.720 & -0.062 & -0.457 & 0.339\\
           & \refsim & 4.650 & -0.112 & -0.360 & 0.353   & 4.124 & -0.111 & -0.327 & 0.353\\
    \wmapa & \agna   & 4.105 & -0.061 & -0.391 & 0.363   & 3.614 & -0.105 & -0.424 & 0.363\\
           & \agnb   & 3.917 & -0.073 & -0.432 & 0.364   & 3.599 & -0.108 & -0.462 & 0.364\\
           & \agnc   & 3.842 & -0.071 & -0.461 & 0.369   & 3.640 & -0.094 & -0.485 & 0.369\\
    \hline
           & \nua    & 4.553 & -0.072 & -0.467 & 0.373   & 4.099 & -0.114 & -0.515 & 0.373\\
           & \nub    & 4.498 & -0.070 & -0.458 & 0.374   & 4.053 & -0.112 & -0.511 & 0.375\\
    \wmapb & \nuc    & 4.411 & -0.074 & -0.449 & 0.373   & 3.985 & -0.114 & -0.504 & 0.375\\
           & \nud    & 4.329 & -0.070 & -0.446 & 0.374   & 3.901 & -0.111 & -0.499 & 0.376\\
           & \nue    & 4.055 & -0.068 & -0.402 & 0.385   & 3.646 & -0.108 & -0.450 & 0.386\\
    \hline
    \multicolumn{2}{l}{\citet{Duffy2008}} & & & &        &  4.11 & -0.084 & -0.47 \\\hline
   \end{tabular}
   \label{tab:DuffyABC}   
  \end{table*}  

\label{lastpage}
\end{document}